\documentclass[10pt,letterpaper]{article}
\usepackage[top=0.85in,left=2.75in,footskip=0.75in]{geometry}

\usepackage{amsmath,amssymb}

\usepackage{changepage}

\usepackage[utf8x]{inputenc}
\usepackage{hyperref}
\usepackage{textcomp,marvosym}

\usepackage{cite}

\usepackage{nameref}

\usepackage[right]{lineno}

\usepackage{microtype}
\DisableLigatures[f]{encoding = *, family = * }

\usepackage[table]{xcolor}

\usepackage{array}

\newcolumntype{+}{!{\vrule width 2pt}}

\raggedright
\usepackage[aboveskip=1pt,labelfont=bf,labelsep=period,justification=raggedright,singlelinecheck=off]{caption}

\usepackage{graphicx}
\usepackage{epstopdf}

\begin{document}
\vspace*{0.2in}
\begin{flushleft}
{\Large
\textbf\newline{Quantifying horizon dependence of asset prices:\\ a cluster entropy approach} 
}
\newline
\\
Linda Ponta\textsuperscript{1},
Anna Carbone\textsuperscript{2},
\\
\bigskip
\textbf{1}  LIUC-Universit\`a Cattaneo, corso Giacomo Matteotti 22, 21053 Castellanza, Italy
\\
\textbf{2} Politecnico di Torino, corso Duca degli Abruzzi 24, 10129 Torino, Italy
\\
\bigskip
\end{flushleft}
\section*{Abstract}
Market dynamic is quantified in terms of the  entropy $S(\tau,n)$  of the clusters formed by the intersections between the series of the prices $p_t$ and the moving average  $\widetilde{p}_{t,n}$. The entropy $S(\tau,n)$ is defined according to Shannon as  $\sum P(\tau,n)\log P(\tau,n)$, with $P(\tau,n)$ the probability for the  cluster to occur with duration $\tau$. 
\par
The investigation  is performed on high-frequency data of the Nasdaq Composite, Dow Jones Industrial Avg and Standard \& Poor 500 indexes downloaded from the Bloomberg terminal. The cluster entropy $S(\tau,n)$  is analysed in raw and sampled data  over a broad range of temporal horizons $M$ varying from one to twelve  months over the year 2018. The cluster entropy $S(\tau,n)$ is integrated over the cluster duration $\tau$  to yield the \emph{Market Dynamic Index} $I(M,n)$, a synthetic figure of price dynamics. A systematic dependence of the cluster entropy $S(\tau,n)$ and the \emph{Market Dynamic Index} $I(M,n)$ on the temporal  horizon $M$ is evidenced. 
\par
Finally, the  \emph{Market Horizon Dependence}, defined as $H(M,n)=I(M,n)-I(1,n)$, is compared with  the horizon dependence of the pricing kernel  with different representative agents obtained via a Kullback-Leibler entropy approach.
  The \emph{Market Horizon Dependence} $H(M,n)$ of the three assets is compared against the values obtained by implementing the  cluster entropy $S(\tau,n)$ approach on artificially generated series (Fractional Brownian Motion).

\section{Introduction}
Entropy, as a tool to quantify heterogeneity and dynamics in complex systems, has found a number of applications in different contexts \cite{crutchfield2012between,bandt2002permutation,grassberger1983characterization,marcon2014generalization,karpiarz2014international,rubido2018entropy}.
In economics and finance, the entropy ability to quantify heterogeneity and disentangle ordered  and disordered patterns in data relevant to complex systems, has been adopted  for portfolio selection to outperform traditional methods based on  Markowitz covariance and Sharpe single-index models \cite{philippatos1972entropy,fernholz2002stochastic,ou2005theory,xu2011portfolio,usta2011mean,zhou2013portfolio,zhang2012possibilistic, bera2008optimal,demiguel2009optimal,rodder2010entropy,
chandrinos2018construction,gospodinov2017general,chen2017study,ormos2014entropy,lahmiri2018informational,lahmiri2018randomness,lahmiri2018long,lahmiri2017disturbances,lahmiri2017nonlinear}.
Entropy ability to quantify dynamics other than heterogeneity has gained interest to the aim of implementing entropy-derived tools  to shed light on fundamental aspects of asset pricing dynamics beyond portfolio optimization \cite{Hansen1991implications,Hansen2014Nobel,Hansen2019Macroeconommic,backus2014sources,ghosh2017what}.
\par
Macroeconomic shocks  are becoming increasingly important due to the growing connectedness of the assets in a global economy. The propagation of these  shocks, which intrinsically are not diversifiable, cannot be averaged out   by diversifying investments and, thus, even the best selection of portfolio assets might fail in keeping investors safe.
Asset pricing models aim at providing estimates of endogenous risk by quantifying market evolution in terms of a stochastic function: the {\em pricing kernel} $m_{t}$. Equilibrium prices $p_t$ of traded securities can be represented as the conditional expectation of the discounted future payoff $z_t$:
\begin{equation}
p_t= E\left [\frac{m_{t+1}}{m_{t}}z_{t+1}\right ] \hspace{15pt},
\end{equation}
where $m_{t+1}/m_{t}$ is known as the {\em stochastic discount factor}. The  pricing kernel $m_{t}$ is  factorizable into a function of the consumption growth $\mu_{t+1}$  times  $\psi_{t}$ (a model specific term):
\begin{equation}
m_t= \mu_{t+1}\psi_{t} \hspace{15pt}.
\end{equation}
The standard consumption-based asset pricing model identifies the pricing kernel as a simple parametric function of the consumption growth ${C_t}$. In this framework,   with time-separable power utility representative agent models, the function $\mu_{t+1}$ is simply proportional to $\Delta C_t = \log ({C_t}/{C_{t-1}})$. More sophisticated agent behaviours have been suggested  to explain puzzling phenomena such as  amplitude and cross-sectional dispersion of returns among different categories of financial assets, equity premia and risk-free rates.
\par
Pricing kernel dispersion and  dynamics with different representative agents are modelled by using the Kullback-Leibler entropy in \cite{backus2014sources} and  extending the findings of \cite{Hansen1991implications}. The work \cite{Hansen1991implications} was addressed to quantify standard deviation and volatility to define the pricing kernels bounds.  A lower bound was provided for the
volatility of the permanent component of asset pricing kernels, showing that stochastic discount factors need to be very volatile to be consistent with high Sharpe ratios  \cite{Hansen1991implications}.
A relative entropy minimization approach, based on the Kullback-Leibler divergence,   is put forward in \cite{ghosh2017what} to extract the model dependent term $\mu_{t+1}$ and quantify the minimum amount of extra information to be embedded in the standard pricing kernel models for reproducing asset return correctly.
The Kullback-Leibler divergence between the probability distribution functions of the components $\mu_{t+1}$ and $\psi_{t}$ has been used as criterion to estimate the deviation of  $m_{t+1}$ with respect to the simple consumption flow growth model. It was argued that the Kullback-Leibler divergence criterion  is equivalent to maximize the  entropy of the fundamental pricing kernel component \cite{ghosh2017what}.
\par
An information theoretical tool has been recently developed which yields  the weights of the efficient portfolio  by using the \emph{cluster entropy} estimated via the detrending moving average algorithm proposed in \cite{carbone2013information,carbone2007scaling,carbone2004analysis}. Interestingly, the works \cite{ponta2017detrending,ponta2018information} show that the cluster entropy of the volatility takes values depending on each market, as opposed to the entropy of the prices, which was shown to be approximately invariant across the markets.   The  \emph{Market Heterogeneity Index}, defined as the integral of the  cluster entropy, provides a cumulative figure allowing a straightforward comparison with the portfolio weights obtained by the Sharpe ratio approach. The main advantage of the cluster entropy approach is not to require  a specific distribution  of returns like a symmetric Gaussian distribution. Such a distribution is quite elusive in real-world financial assets and thus hindering, in principle, the application of Markowitz-based portfolio models.
\par
In this work, we implement the cluster entropy approach for quantifying the intrinsic dynamics  of prices  and capturing the endogenous sources of risk over different temporal horizons.
The present work  builds upon and extends the study \cite{ponta2018information} that was limited  to extract the portfolio weights from the cluster entropy of the prices and volatility of the financial series over a constant time horizon  (about 6 years from 1998 to 2004). Under the condition of constant temporal horizon, the cluster entropy of the prices has been found to be almost invariant across the markets in \cite{ponta2018information}. 
\par
Here the focus is on gaining insights in the intrinsic dynamic ruling price evolution. Hence, the cluster entropy analysis is performed over multiple horizons. The horizon dependence was not studied in  \cite{ponta2018information} that reported the quantitative comparison of cluster entropy observed in several markets over the same horizon  (i.e. same time interval of six years 1998-2004).
\par
The ability of the cluster entropy approach to quantify the intrinsic dynamic of the prices is proved by analysing several assets. For the sake of simplicity in this work we report the results obtained on the three markets described  in Table \ref{tab:data}.
\par
Cluster entropy has been analysed for prices of market indices (tick-by-tick prices from Jan $1^{st}$ to  Dec. $31^{st}$ 2018) NASDAQ, DJIA and S$\&$P500 with length $N=6982017$, $N=5749145$ and $N=6142443$ respectively.  Data have been downloaded from the terminal  www.bloomberg.com/professional. The three financial markets have been selected based on homogeneity and similarity criteria. The three markets are traded in the same country and with same currency.  Furthermore the assets are characterised by a comparable number of transactions over time. The similarity criteria rule out differences in the dynamics that might be due to external causes. Another condition ensuring that the observed behavior is genuinely related to the intrinsic price dynamics rather than exogenous factors is to keep limited  the maximum extension of the temporal horizon. In the current study the max temporal extension is one year (12 Months) and the analysis has been performed on multiples of monthly subsets from one to twelve months.  
\par
 It is worth noting that though in literature many study have been performed to understand the asset pricing dynamics using low-frequency data, for example, to estimate the low-frequency components of returns, in this paper the analysis is performed on high-frequency data to investigate and capture the endogenous sources of risk. The data ranges over one single year. The huge data sets allow one to apply the cluster entropy algorithm over monthly segmented series with average lengths of the order of $\sim 500000$.
\par
A systematic dependence of the cluster entropy of the asset prices over varying temporal horizons has been observed, that could be related to the macroeconomic fundamental properties and exogenous dynamics  rather than to simple variations across different markets.
\par
The manuscript is organized as follows. The main relationships relevant to understanding and implementing the cluster entropy approach are shortly recalled in Section \ref{Method}. The  analysed data sets  (financial assets and artificially generated series) are described in Section \ref{Data}. In Section \ref{Results} the cluster entropy and the Market Dynamic Index of the prices series as a function of the temporal horizon $M$ are reported together with a comparison against the Kullback-Leibler entropy results obtained by simulating the pricing kernel with different representative agent models.
The cluster entropy and the market dynamic index estimated for Fractional Brownian Motion (FBM) sequences are reported and discussed. The artificially generated  FBM data are taken as reference to validate the accuracy of the deviations observed in the real-world assets markets and validate the findings via a standard T-paired test. 

\section{Methods}
\label{Method}
In this section,  we briefly recall the main definitions and equations used which are the core computational ingredients of the algorithm. \par
 The cluster entropy  is obtained by taking the intersection of the asset prices $p_t$ and its moving average $\tilde{p}_{t,n}$  for different moving average window $n$  \cite{ponta2017detrending,ponta2018information,carbone2013information,carbone2007scaling,carbone2004analysis}. For each  window $n$, the subsets $\{p_t: t=s,....,s-n \}$  between two consecutive intersections are considered. The subsets are named  \emph{clusters}.   The clusters are exactly defined as the portions of the series between death/golden crosses according to the technical trading rules. Therefore, the information content has a straightforward connection with the trader's perspective on the price and volatility series. Then, the clusters are ranked according to their characteristic size, the duration $\tau$. The probability distribution function $P(\tau,n)$ of the cluster duration is obtained.
The present  approach directly yields either power-law or exponential distributed cluster distributions, thus enabling us to separate the sets of inherently correlated/uncorrelated blocks along the sequence.
The continuously compounded return is defined  by:
\begin{equation}
\label{returnlin}
 r_t = p_t - p_{t-h} \hspace{10pt} ,
\end{equation}
where $p_t$ is the price at the time $t$,  with $ 0<h<t<N $ and $N$ the maximum length of the time series.
Alternatively, one can consider the log-return  defined as:
\begin{equation}
\label{returnlog}
r_t  = \log p_t - \log p_{t-h} \hspace{10pt}.
\end{equation}
\par
The approach adopted in  this work builds upon the idea of Claude Shannon to quantify the ‘expected’  information contained in a message extracted from a sequence $\{x_t \}$ \cite{shannon1948mathematical} by using the entropy functional:
\begin{equation}
S[P(x_t)] = \int_X p(x_t) \log p(x_t) dx_t \hspace{5pt},
\label{Shannon_int}
\end{equation}
with $P$ a probability distribution function associated with the sequence $\{x_t \}$.
For discrete sets, Eq. (\ref{Shannon_int})  reduces to:
\begin{equation}
S[P(x_t)] = \sum_X p(x_t) \log p(x_t) \hspace{5pt}.
\label{Shannon}
\end{equation}
Consider the time series  $\{x_t \}$ of length $N$ and the moving average $\{\widetilde{x}_{t,n}\}$   of length $N-n$  with $n$ the moving average window.
The function $\{\widetilde{x}_{t,n}\}$ generates, for each $n$, a partition  $\{\cal{C}\}$  of non-overlapping clusters between two consecutive intersections of $\{x_t \}$ and
$\{\widetilde{x}_{t,n}\}$. Each cluster $j$ has  duration:
\begin{equation}
\label{l} \tau_j\equiv  \|t_{j}-t_{j-1}\|
\end{equation}
\noindent
where the instances $t_{j-1}$ and $t_j$ refer to two subsequent  intersections.
 The probability distribution function  $P(\tau,n)$  can be obtained by ranking the number of clusters ${\mathcal N}(\tau_1,n),{\mathcal N}(\tau_2,n), ..., {\mathcal N}(\tau_j,n)$ according to their length $\tau_1, \tau_2,..., \tau_j$ for each $n$. A stationary sequence of clusters $\cal{C}$ is generated with  probability distribution function varying as \cite{carbone2013information}:
\begin{equation}
\label{Pl} P(\tau,n)\sim\tau^{-\alpha} {\mathcal F}\left({\tau},{n}\right) \hspace{5pt},
\end{equation}
with the factor ${\mathcal F}\left({\tau},{n}\right)$ taking the form $ \exp({-\tau}/{n})$,  to account for the finite size effects when $\tau\gg n$, resulting in  the  drop-off of the power-law and the onset of the exponential decay.
The cluster entropy  writes (the details of the derivation can be found in \cite{carbone2013information,carbone2004analysis}):
\begin{equation}
S[P(\tau_j,n)] = \sum_j P(\tau_j,n)\log P(\tau_j,n) \hspace{5pt},
\label{Shannon}
\end{equation}
that by using Eq.~(\ref{Pl}) simplifies to:
\begin{equation}
\label{lentropy2}
S(\tau,n)=S_0+\log\tau^{\alpha}+{\tau\over n}\hspace{5pt},
\end{equation}
where $S_0$ is a constant, $\log\tau^{\alpha}$ and $\tau/ n$  are related respectively to the terms $\tau^{-{\alpha}}$ and ${\mathcal F}(\tau,n)$.
The minimum value of the entropy is obtained for the fully ordered (deterministic) set of clusters with  duration $\tau=1$. Eq.~(\ref{lentropy2})  in the limit $n\sim\tau\rightarrow1$ and  $S_0\rightarrow-1$   reduces to $S(\tau,n)\rightarrow0$. Conversely, the maximum value of  the entropy $S(\tau,n)=\log N^{\alpha}$ is obtained when $n\sim\tau\rightarrow N$ (with $N$  the maximum length of the sequence). This condition corresponds to the maximum randomness (minimum information) carried by the sequence,  when a single longest cluster is obtained coinciding with the whole series. 
\par
For a fractional Brownian motion, the exponent $\alpha$ is equal to the fractal dimension $D=2-H$  with $H$ the Hurst exponent of the time series. The term $\log\tau^{\alpha}$ can be thus interpreted  as a generalized form of the Boltzmann entropy $S=\log\Omega$, where $\Omega = \tau^D$ corresponds to the fractional volume occupied by the fractional random walker.
The term $\tau/n$ represents  an excess entropy (excess noise) added to the intrinsic entropy term $\log\tau^D$ by the partition process. It depends on $n$ and is related to the finite size effect discussed above.
\par
We stress the difference between the time series partitions  obtained   either by using equal size boxes or moving average clusters.
For equal size boxes, the excess noise term ${\tau / n}$  vanishes (as it becomes a constant that can be included in the constant term) thus the entropy reduces to the logarithmic term as found in Ref.~\cite{grassberger1983characterization}, which corresponds to the intrinsic entropy of an ideal fractional random walk. When a moving average partition is used, an excess entropy term  ${\tau / n}$  emerges accounting for the additional heterogeneity introduced by the random partitioning process operated by the moving average intersections.
\par
To univocally quantify  market properties through the entropy Eq.~(\ref{lentropy2}), a cumulative information measure has been defined as follows:
\begin{equation}
I(n)=\int_0^{\tau_{max}} S (\tau,n)d\tau \hspace*{5 pt},
\label{Integral}
\end{equation}
which, for discrete sets, reduces to:
\begin{equation}
I(n) = \sum_{\tau} S (\tau,n) \hspace{5pt}.
\label{Integrald}
\end{equation}
\par
 The function $I(n)$ has been used to quantify cross-market heterogeneity in \cite{ponta2018information}. The  cluster entropy of the volatility $v_T$ was integrated over the cluster duration $\tau$  to the purpose of obtaining the  weights of the optimal portfolio.
\par
In this work, the function $I(n)$ will be used to quantify the intrinsic market dynamic. The cluster entropy of the prices will be integrated over the cluster duration $\tau$  to the purpose of obtaining the horizon dependence.
\par
As a concluding remark to this section, it is worth mentioning the relation between the cluster entropy  approach adopted in this work, the  multiscale entropy (MSE) and its variants \cite{costa2002multiscale,niu2015quantifying,humeau2015multiscale}. 
The multiscale entropy provides insights into the complexity of fluctuations over a range of time scales  and thus extends the standard one-sample entropy. 
The computational implementation of multiscale entropy implies a coarse graining of the time series at increasingly time resolutions. Coarse graining the data basically means averaging different numbers of consecutive points to create different scales or resolutions of the signal. In the cluster entropy approach proposed here, the coarse graining of the signal is performed through the moving average, i.e. a time dependent averaging.
The multiscale entropy analysis aims at quantifying the interdependence between entropy and scale, achieved by evaluating sample entropy of univariate time series coarse grained at multiple temporal scales. This facilitates the assessment of the dynamical complexity of the system whose behavior is reflected by the time series data.

\section{Data}
\label{Data}
Prices of market indices traded in the US, namely NASDAQ, DJIA and S$\&$P500 are investigated.
Data sets have been downloaded from the terminal  www.bloomberg.com/professional.
For each index, the data set includes tick-by-tick prices $p_t$ from January to  December 2018.
Details (Ticker; Extended name; Country; Currency; Members; Length) as provided by Bloomberg for the three assets are reported in Table \ref{tab:data}. The length of each index is referred to the year 2018 (last column).
Different temporal horizons have been considered as monthly integer multiples of one-month period $M$ ranging from $M=1$ up to $M=12$.  The individual lengths of the subsequences referred to the twelve time periods  are reported for each index in Table \ref{tab:sampleddata}.
\par
To the purpose of performing cluster entropy analysis over sequences with constant lengths, the raw data are sampled, thus yielding data series with equal length.
The sampling frequency  is defined for each series by dividing the length of the series corresponding to the longest horizon  by the minimum, rounding the ratio to the nearest whole,  that is used to sample the raw data. 
\par
Consider for example the S$\&$P500 market ($3^{rd}$ column  in Table \ref{tab:sampleddata}). The minimum value  of the length is that at $M=1$  (January with $N=516635$) and the maximum value is longest horizon of interest (for example $N=5180006$ for horizon $M=10$ equal to ten months from  January to October). As the sampling frequency is different for each series we consider the minimum value to perform the analysis with the same length.
\par
Furthermore for the sake of validating the obtained results, a set of computational tests have been performed on artificially generated series of different lengths $N$. The artificial series  have been generated by means of the FRACLAB tool available at:  https://project.inria.fr/fraclab/. The artificial series have been generated with lengths $N$ corresponding to those of the financial markets under investigation (Table \ref{tab:sampleddata}). Further details and results are reported in the following Sections.


\section{Results}
\label{Results}
\par
Probability distribution  $P(\tau,n)$ and entropy $S(\tau,n)$ have been calculated for a large set of prices series  by means of the procedure summarized in Section \ref{Method}.  The series of the NASDAQ, DJIA and S$\&$P500 indexes described in Section \ref{Data}  have been used for the investigation.
\par
\begin{figure}
\includegraphics[scale=0.3]{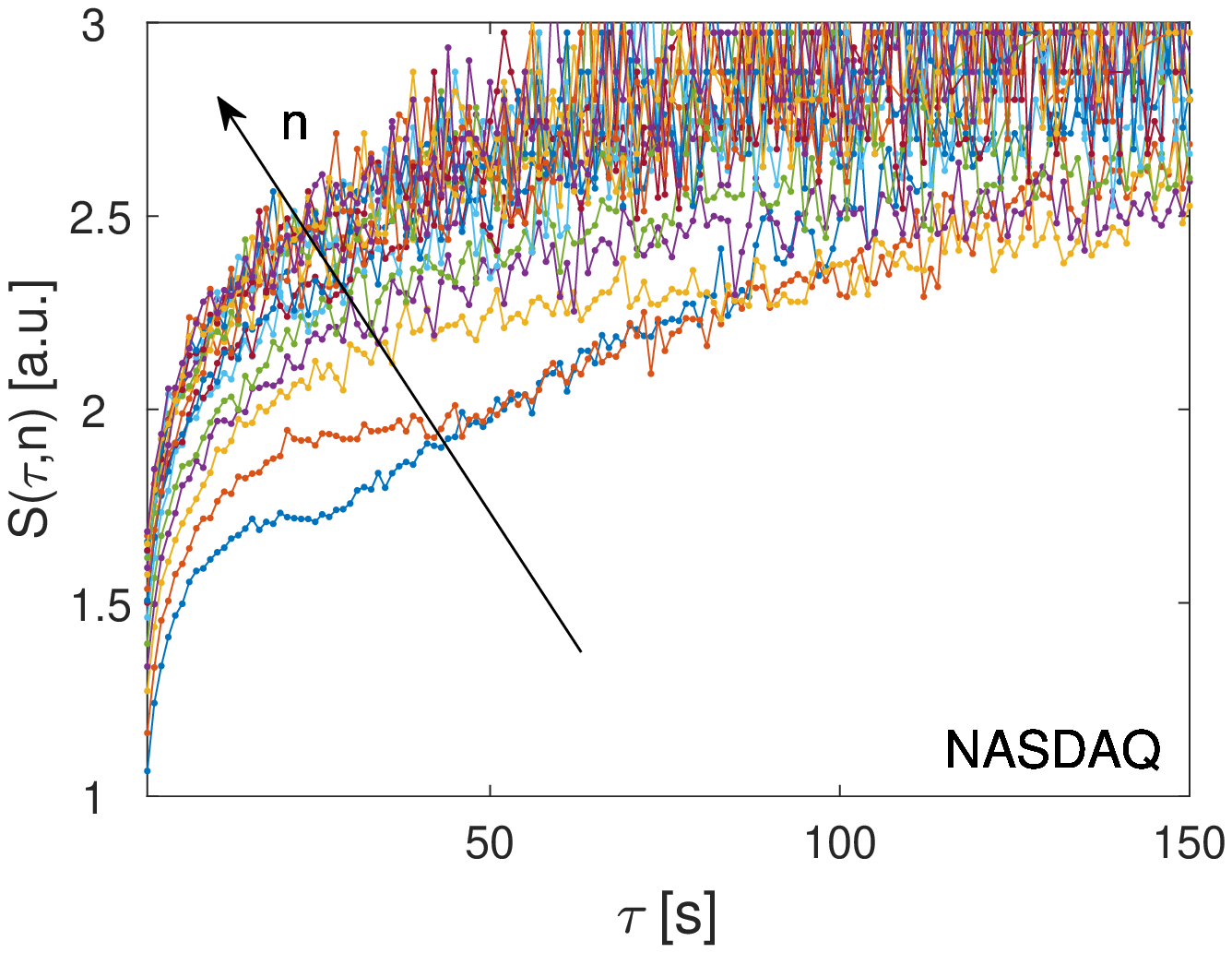}
\includegraphics[scale=0.3]{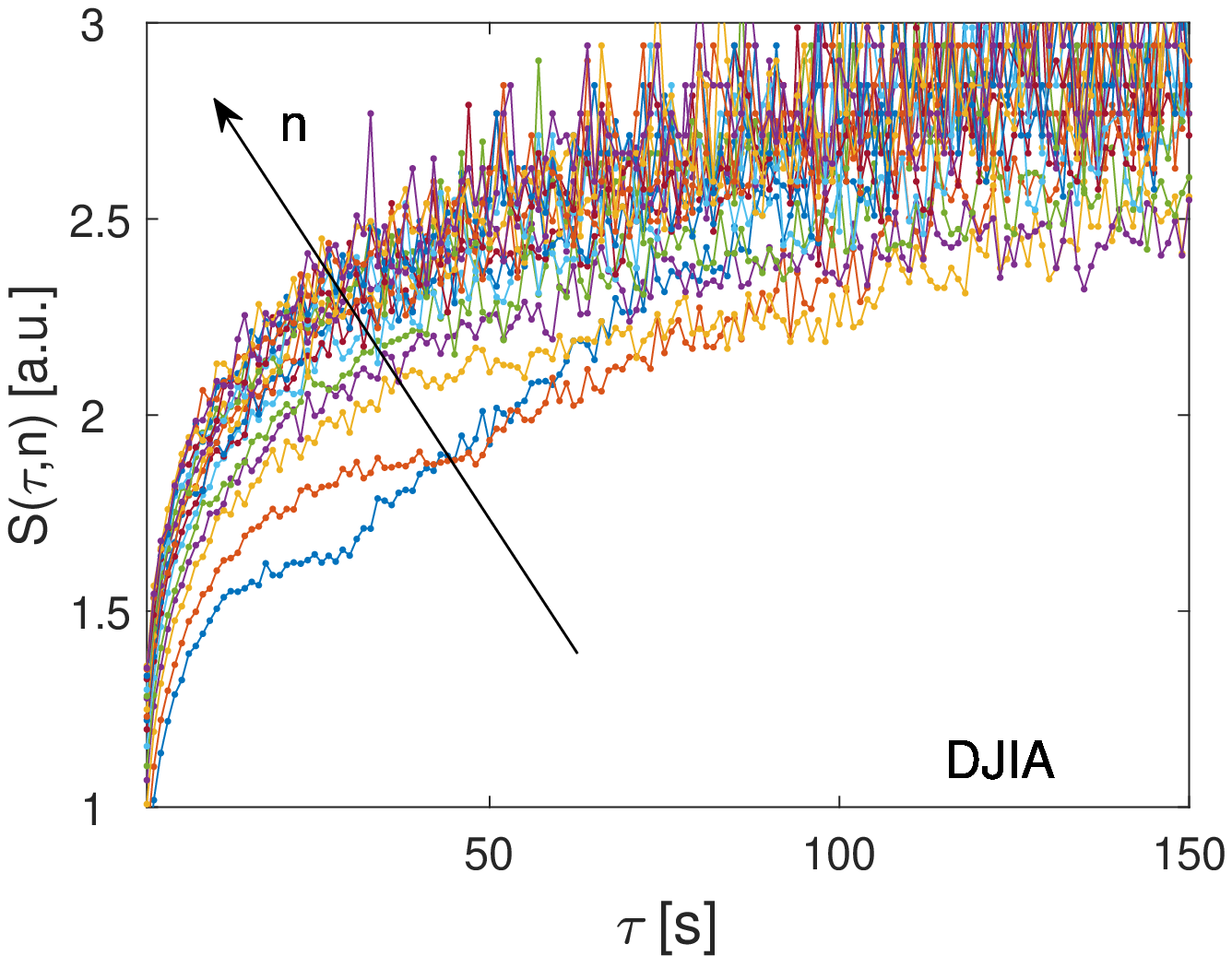}
\includegraphics[scale=0.3]{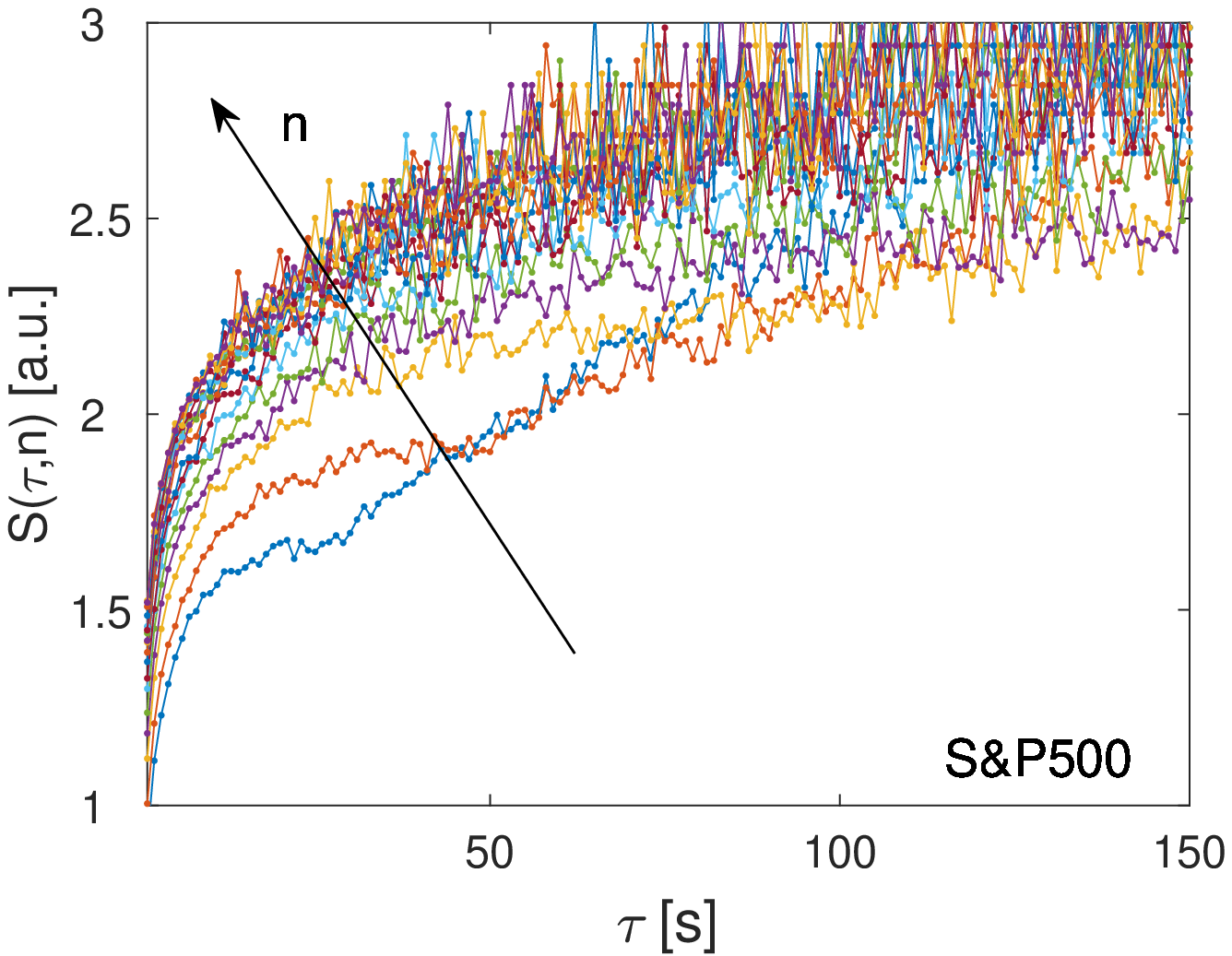}
   \caption{
Cluster entropy $S(\tau,n)$  vs duration $\tau$ for the time series  of the prices (raw data) respectively of the market indices NASDAQ, DJIA and S$\&$P500 described in Table \ref{tab:data}. The series lengths  are  $N=586866$, $N=516644$ and $N=516635$  respectively for NASDAQ, DJIA and S$\&$P500 as given in Table \ref{tab:sampleddata}. The curves refer to one period, i.e. the first month of tick-by-tick data ($M=1$). Different curves in each figure refer to different values of the moving average window $n$ as indicated by the arrow. }
 \label{Fig:entropyrawpriceM1}
\end{figure}

Fig.~\ref{Fig:entropyrawpriceM1}   shows the cluster entropy $S(\tau,n)$ calculated by using raw data prices. In particular, the plots refer to one month of data ($M=1$). The series lengths  are  $N=586866$, $N=516644$ and $N=516635$  respectively for NASDAQ, DJIA and S$\&$P500 as given in Table \ref{tab:sampleddata}.
\par
\begin{figure}
\includegraphics[scale=0.3]{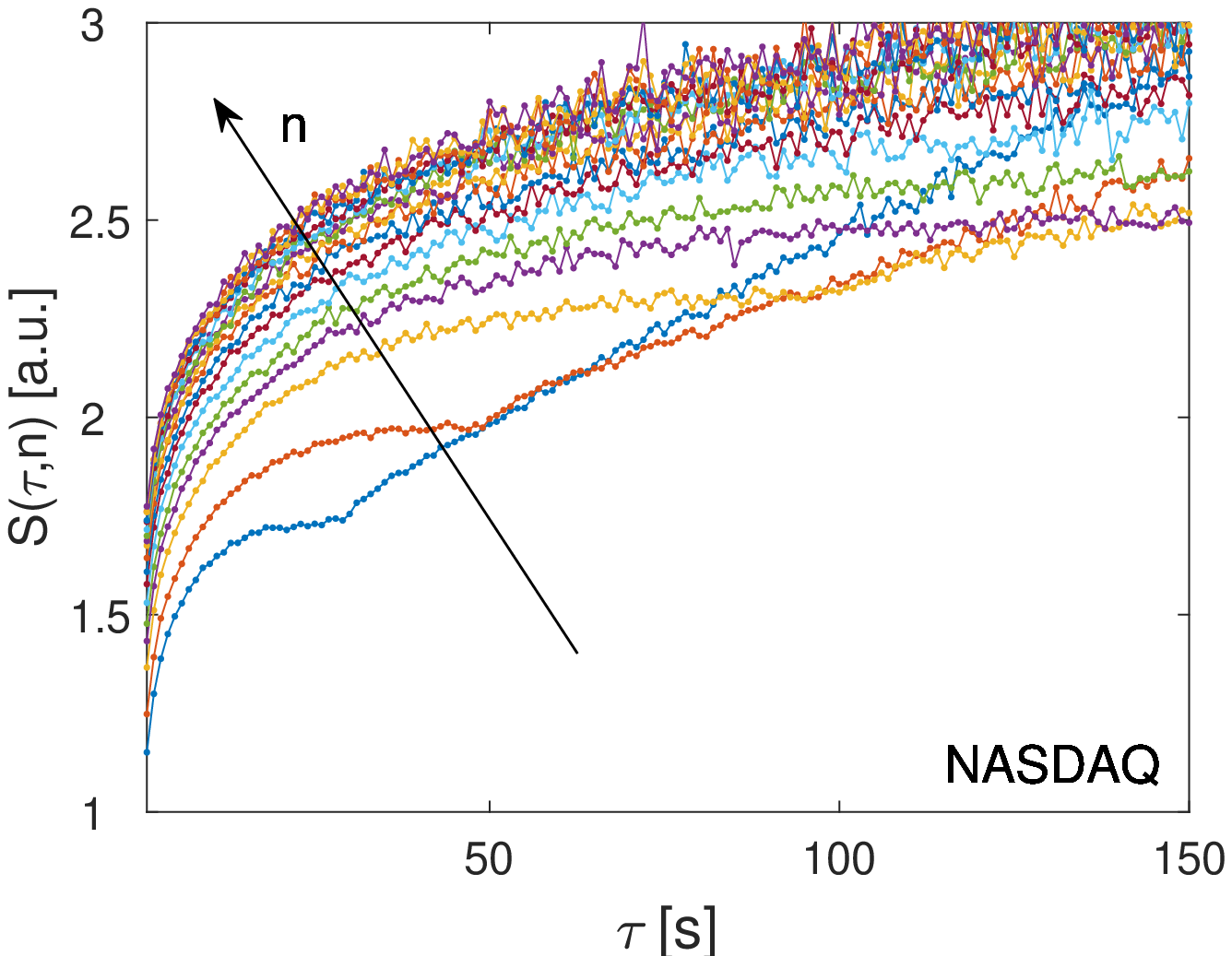}
\includegraphics[scale=0.3]{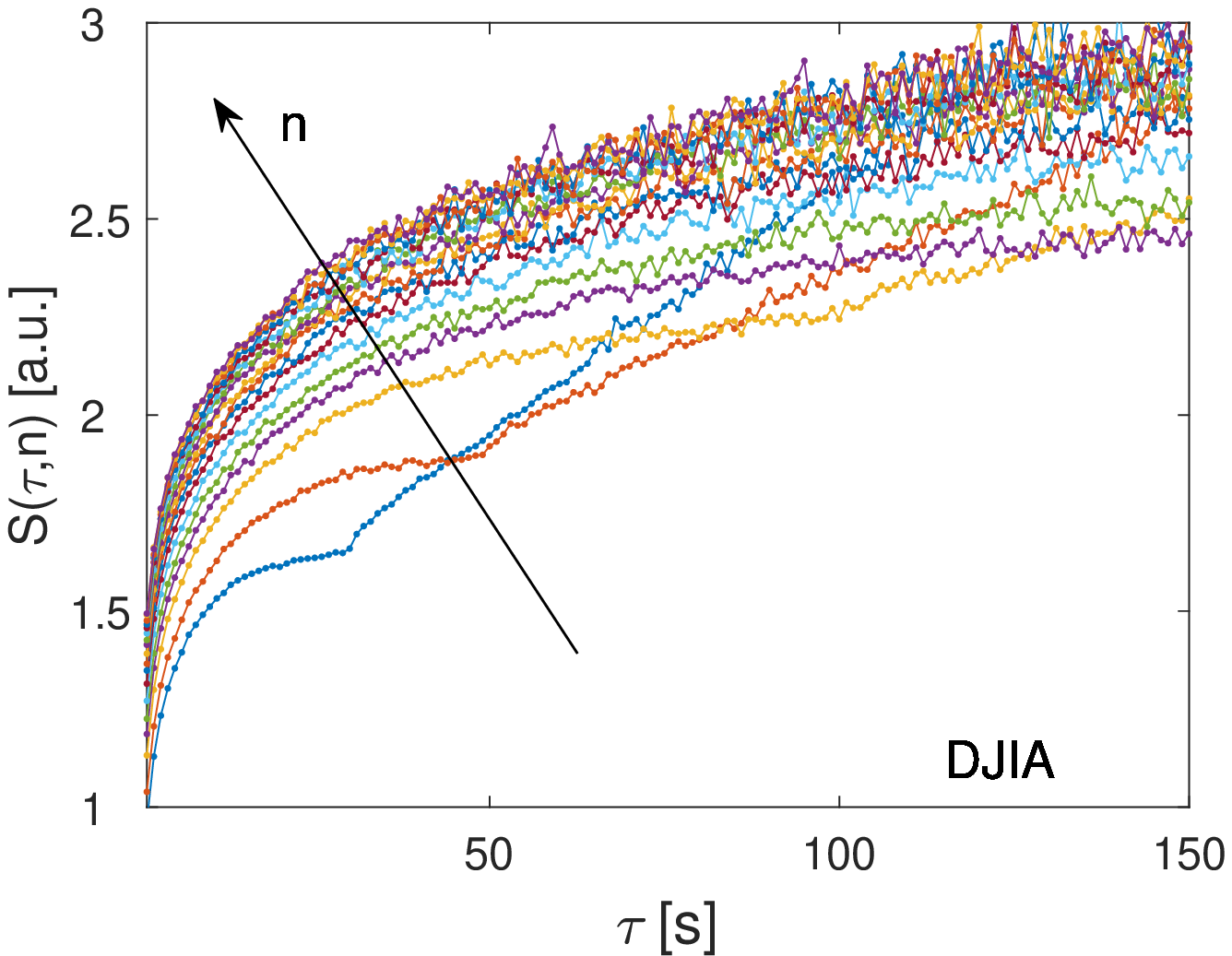}
\includegraphics[scale=0.3]{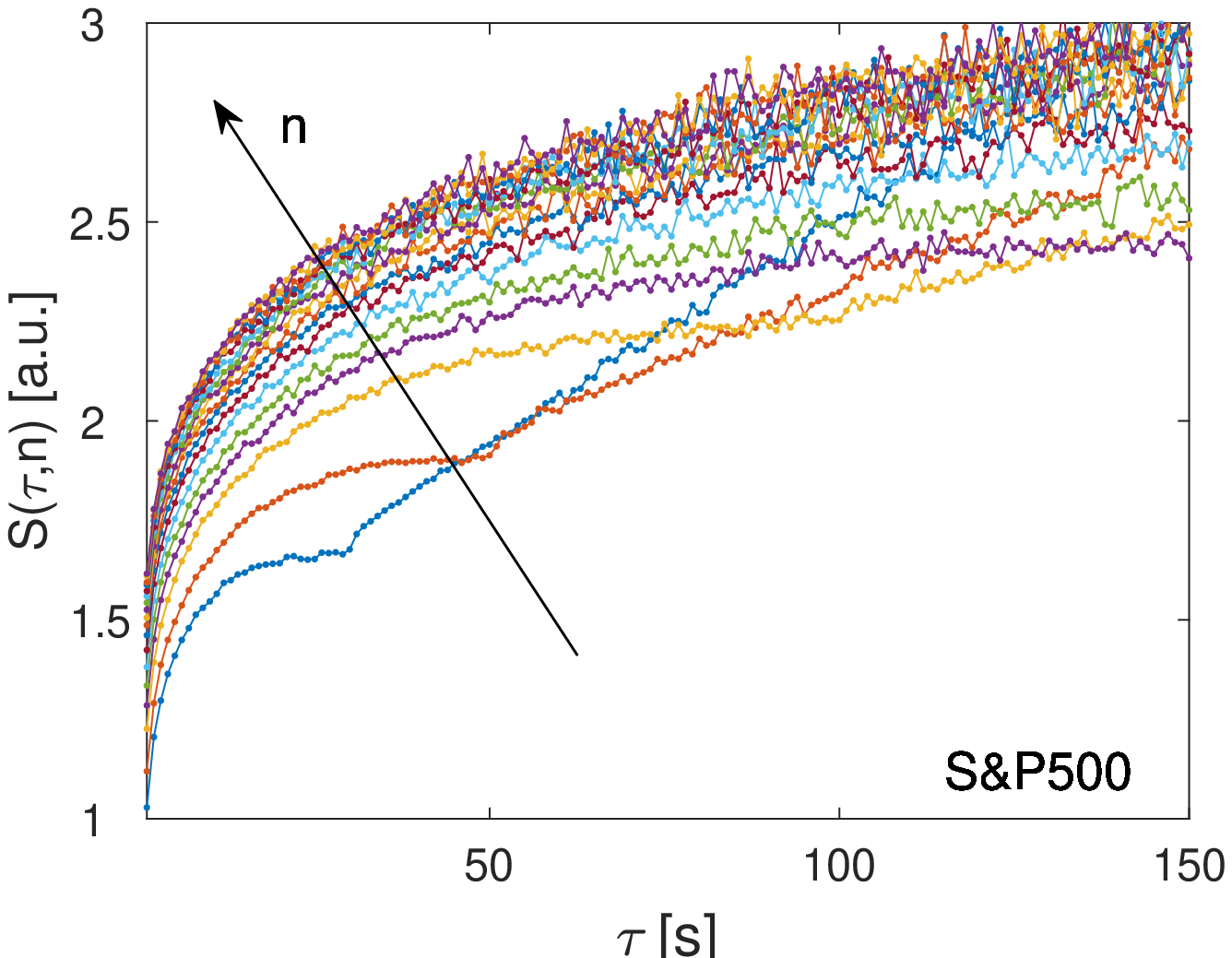}
   \caption{
   Cluster entropy $S(\tau,n)$   vs duration $\tau$ for  the time series  of the prices (raw data)  respectively of the market indices NASDAQ, DJIA and S$\&$P500 described in Table \ref{tab:data}. The series lengths  are  $N= 6982017$, $N=5749145$ and $N=6142443$  respectively for NASDAQ, DJIA and S$\&$P500 as given in Table \ref{tab:sampleddata}. The curves refer to twelve periods, i.e. the whole year 2018 of tick-by-tick data ($M=12$). Different curves in each figure refer to different values of the moving average window $n$ as indicated by the arrow.}
\label{Fig:entropyrawpriceM12}
\end{figure}

Fig.~\ref{Fig:entropyrawpriceM12}   shows the cluster entropy $S(\tau,n)$ calculated by using raw data prices, as in Fig.~\ref{Fig:entropyrawpriceM1}, but here the series refer to a horizon of twelve months  ($M=12$).  The series lengths  are  $N= 6982017$, $N=5749145$ and $N=6142443$  respectively for NASDAQ, DJIA and S$\&$P500 as one can find in the last row of Table \ref{tab:sampleddata}.
\par
\begin{figure}
\includegraphics[scale=0.3]{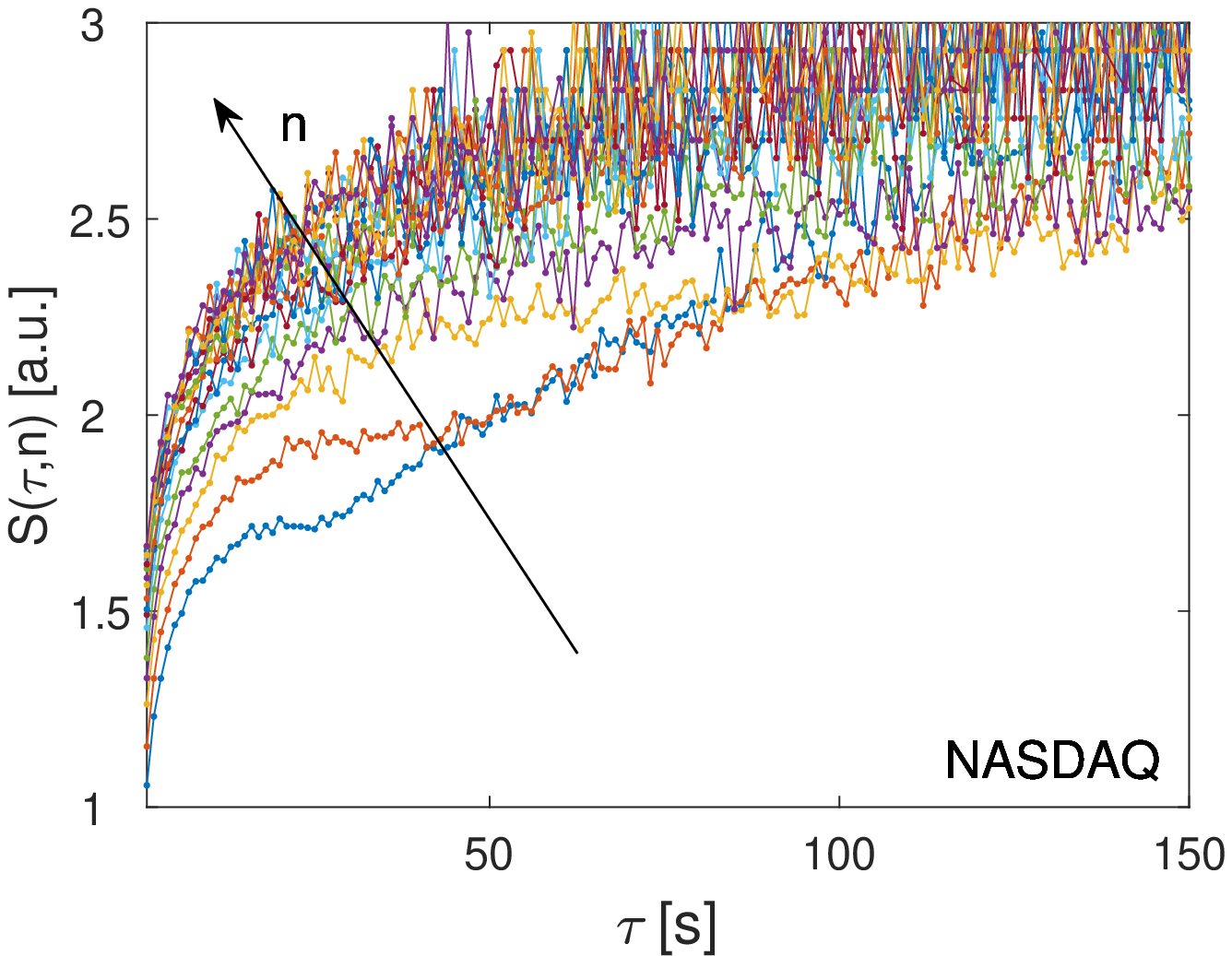}
\includegraphics[scale=0.3]{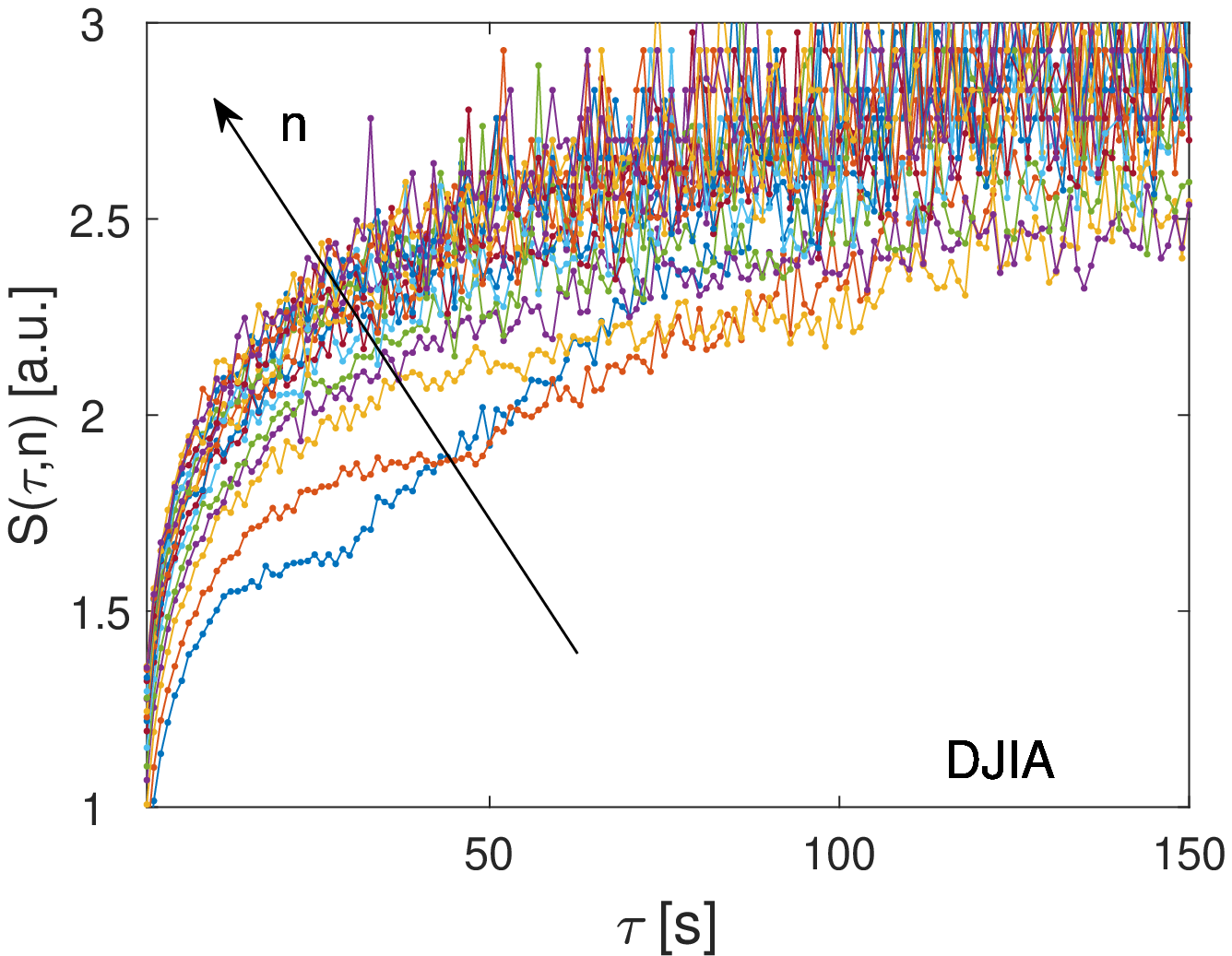}
\includegraphics[scale=0.3]{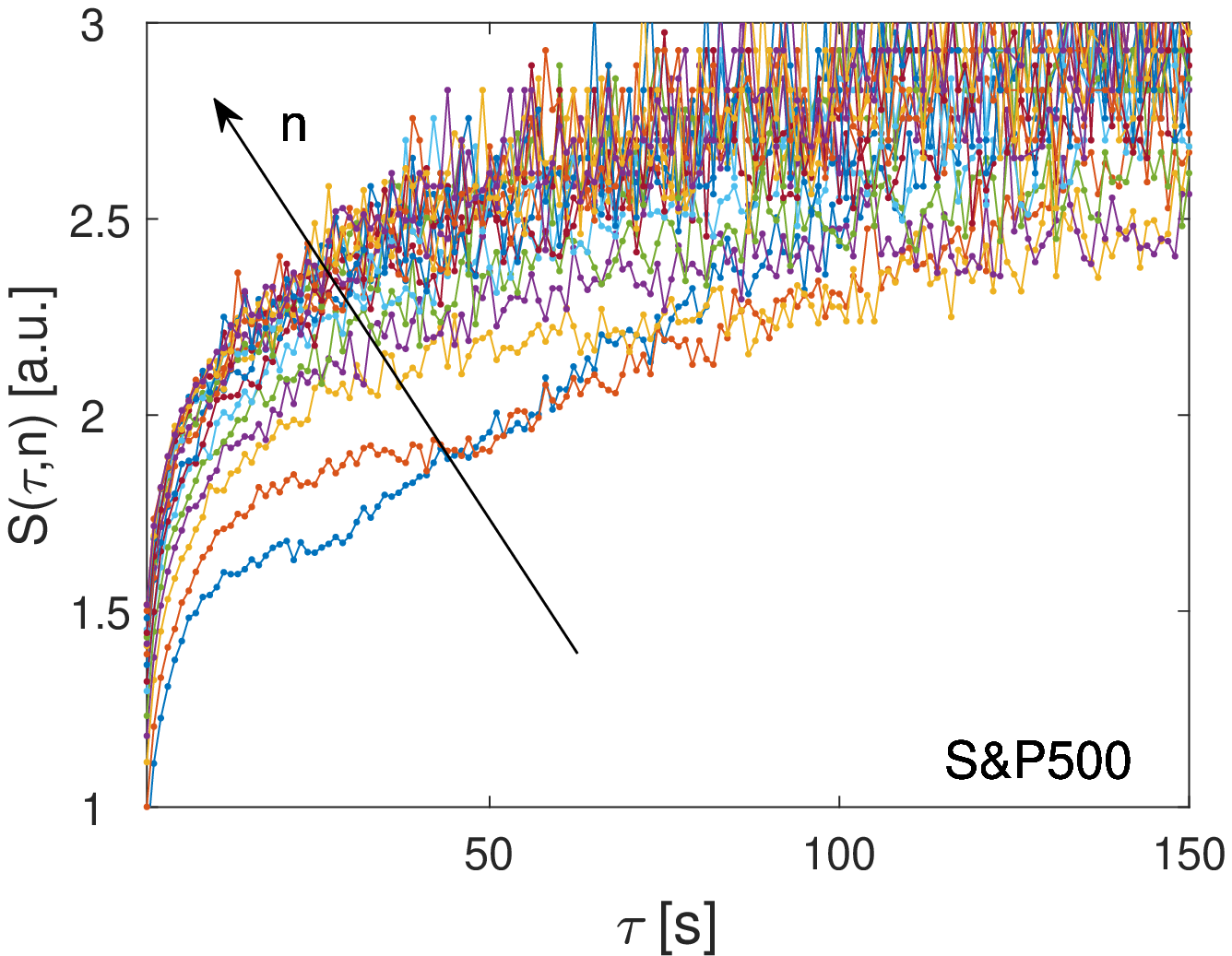}
   \caption{
   Cluster entropy $S(\tau,n)$  plotted   versus cluster duration $\tau$ for the time series of the prices (sampled data) respectively of the market indices NASDAQ, DJIA and S$\&$P500  described in Table \ref{tab:data}.  Figures refer to the first month of data ($M=1$). All time series have same length $N=492035$ obtained by a suitable sampling frequency.
   Different curves refer to different values of the moving average window $n$.}
 \label{Fig:entropysampledpriceM1}
\end{figure}

Fig.~\ref{Fig:entropysampledpriceM1} shows the cluster entropy $S(\tau,n)$ calculated by using the  prices series of the sampled data. The plots refer to the first month of data ($M=1$). All the series have same length $N=492035$.
\par
\begin{figure}
\includegraphics[scale=0.3]{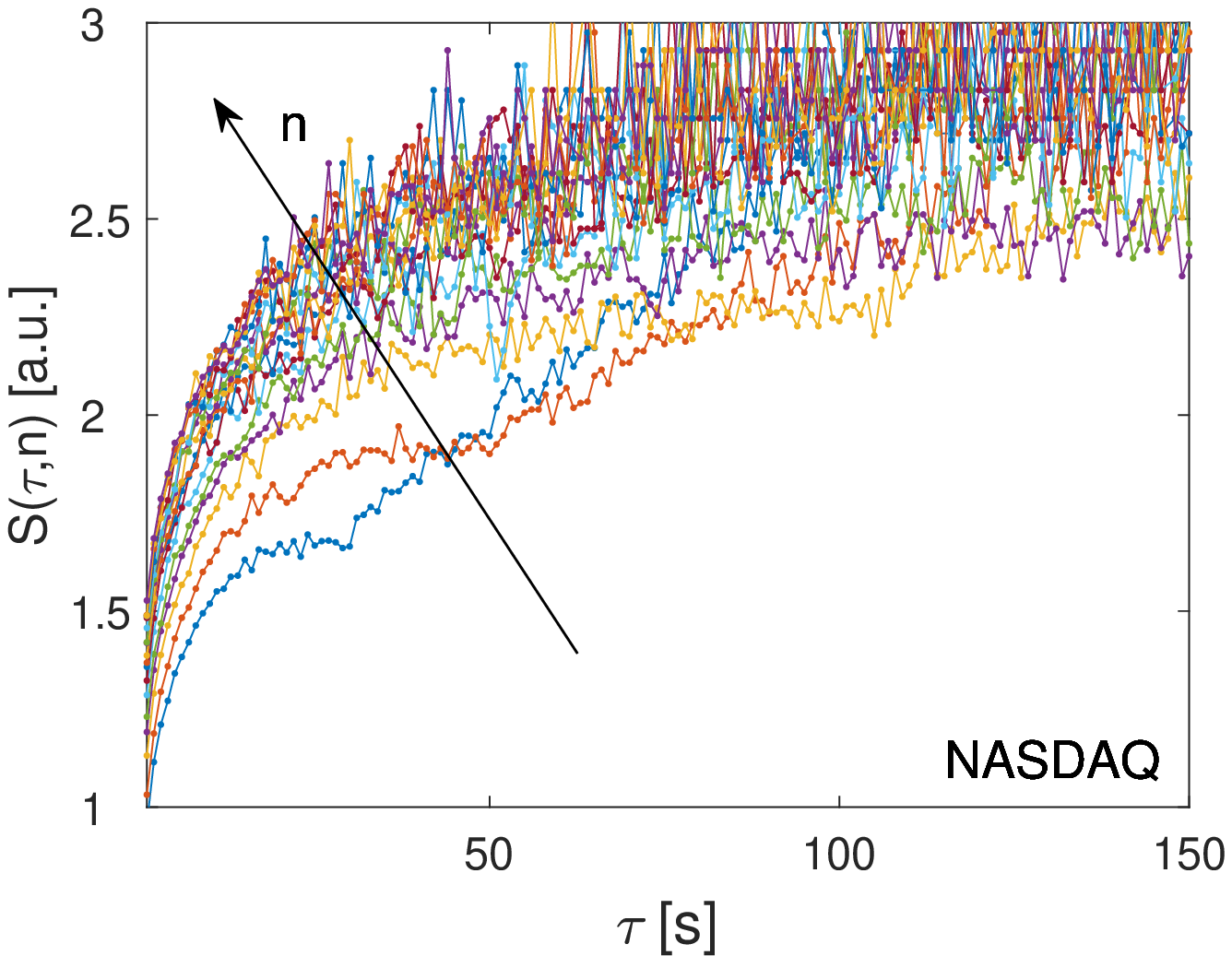}
\includegraphics[scale=0.3]{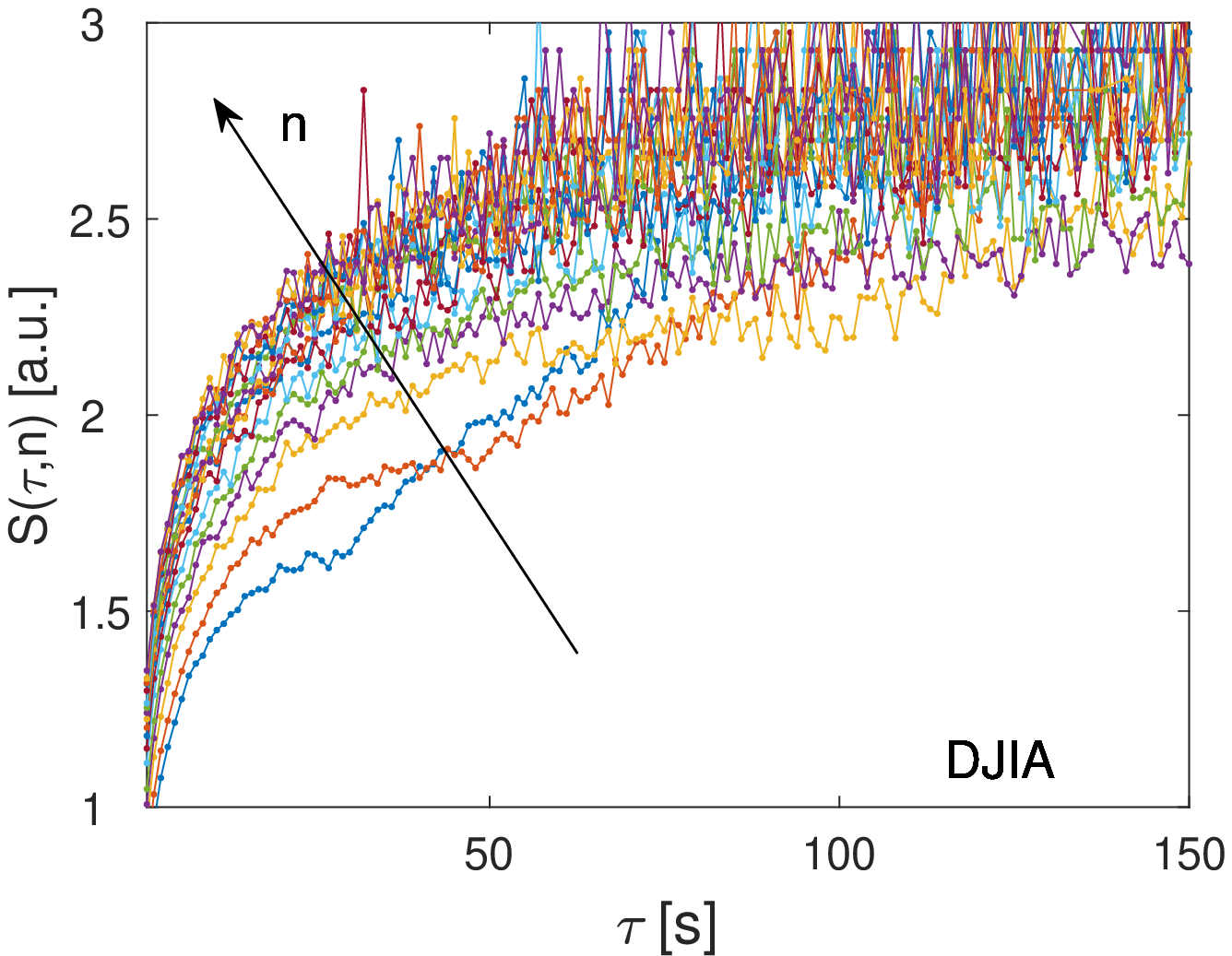}
\includegraphics[scale=0.3]{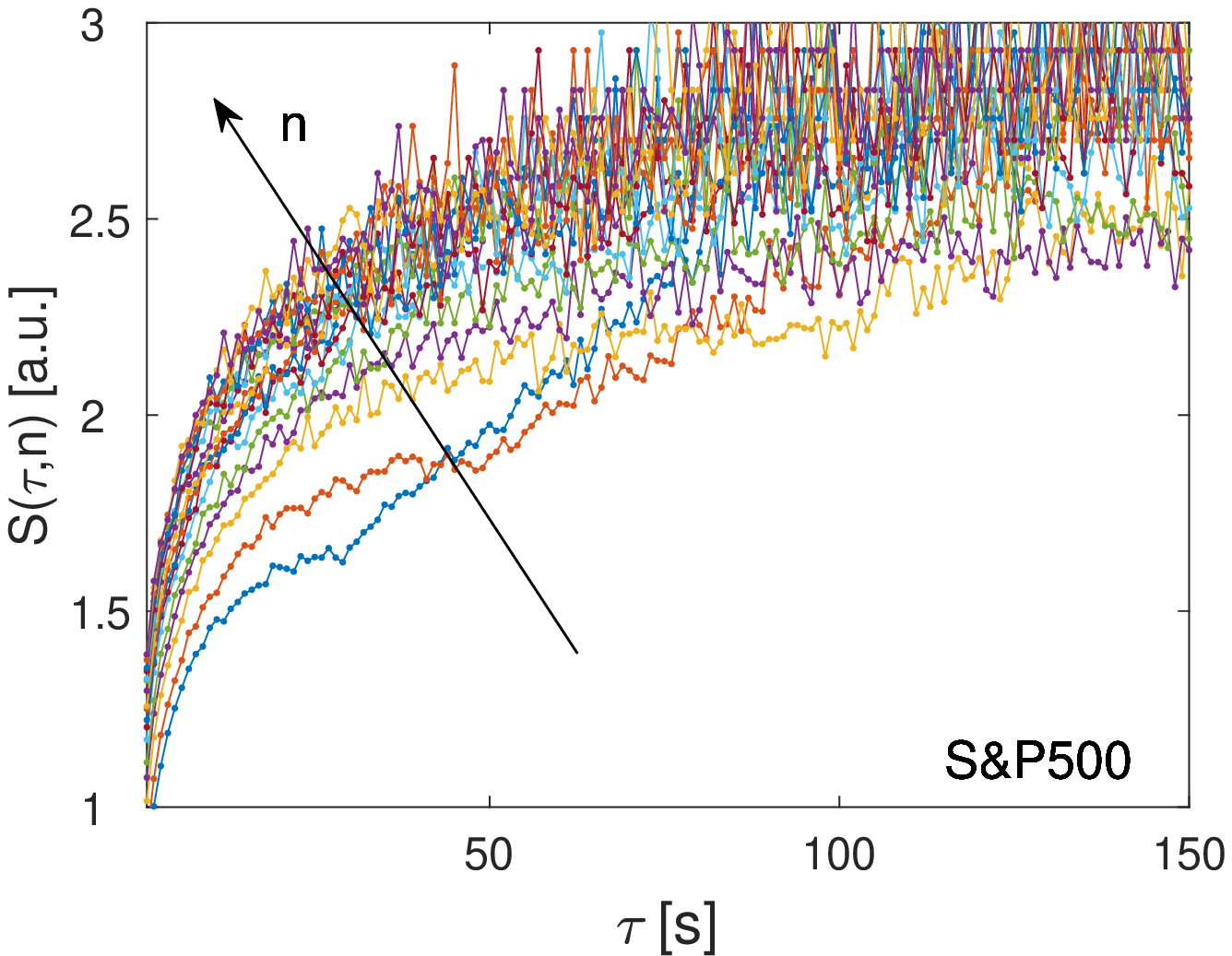}
   \caption{
   Cluster entropy $S(\tau,n)$  plotted   versus cluster duration $\tau$ for the time series of the prices (sampled data) respectively of the market indices NASDAQ, DJIA and S$\&$P500  described in Table \ref{tab:data}.  Figures refer to twelve months of data ($M=12$). All time series have same length $N=492035$ obtained by a suitable sampling frequency.
   Different curves refer to different values of the moving average window $n$.
      }
 \label{Fig:entropysampledpriceM12}
\end{figure}

 Fig.~\ref{Fig:entropysampledpriceM12} shows the cluster entropy $S(\tau,n)$ calculated by using the  prices series of the sampled data. The plots refer to twelve months ($M=12$). All the series  have same length $N=492035$.

\par
Different curves in each figure correspond to moving average values  varying from $n=30\hspace{2pt}\mathrm{s}$,  $n=50\hspace{2pt}\mathrm{s}$,  $n=100\hspace{2pt}\mathrm{s}$,  $n=150\hspace{2pt}\mathrm{s}$,  $n=200\hspace{2pt}\mathrm{s}$ $\ldots$ up to $n=1500\hspace{2pt}\mathrm{s}$ (with step $100\mathrm{s}$).
\par
One can note that the entropy curves exhibit a behaviour consistent with  Eq.~(\ref{lentropy2}). At small values of the cluster duration $\tau \leq n$, entropy behaves as a logarithmic function.  At large values of the cluster duration $\tau \geq n$ the curves  increase  linearly with the term ${\tau /n}$ dominating.
${S}(\tau,n)$  is $n$-invariant for small values of $\tau$, while its slope decreases as $1/n$  at larger $\tau$, as expected according to Eq.~(\ref{lentropy2}), meaning that clusters with duration $\tau > n $
are not power-law correlated, due to the finite-size
effects introduced by the partition with window $n$. Hence, they are characterized
by a value of the entropy exceeding the curve $\log \tau^D$, which corresponds to power-law correlated clusters.  It is worthy to remark that clusters with same duration $\tau$
  can be generated by different values of the moving average window $n$.
At a constant value of $\tau$, larger entropy values are obtained as $n$ increases.
\par
The  entropy ${S}(\tau,n)$ of the  NASDAQ, DJIA and S$\&$P500  prices (shown in Fig.~\ref{Fig:entropyrawpriceM1}, Fig.~\ref{Fig:entropyrawpriceM12}, Fig.~\ref{Fig:entropysampledpriceM1}  and Fig.~\ref{Fig:entropysampledpriceM12})  is representative of a quite general behaviour observed in several markets analysed by using the proposed cluster entropy approach.
\bigskip
\par
In the following, we will discuss how to quantify the horizon dependence  of the asset prices by using the cluster entropy  function $S (\tau,n)$ estimated over different periods $M$. To this purpose, we use the \emph{cumulative information measure}  function  defined in Eq.~(\ref{Integral}).
\par
The quantity $I(M,n)$ is calculated by using the values of the entropy  $S (\tau,n)$  of the asset prices $p_t$ estimated over several periods $M$, ranging from one to twelve months, by using raw and sampled data.
The first period ($M=1$) of the price sequences is taken in correspondence of January 2018  for all the assets. Multiple period sequences have been built by considering  $M=2$ (January and February 2018) and, so on, up to  $M=12$ (one year from January to December 2018).  Details concerning lengths of the series corresponding to the temporal horizons $M$ are reported in Table \ref{tab:sampleddata}.
\par
The \emph{cumulative information measure} $I(M,n)$ has been plotted in Fig.~\ref{Fig:integral} for the prices of the NASDAQ, DJIA and S$\&$P500. One can observe a dependence of the function $I(M,n)$ at different $M$ horizons.
\par
$I(M,n)$ is the same for all $M$ implying that the horizon dependence $H(M,n)$ is negligible at small scales  (small $n$/ small $\tau$ values). Conversely, at large $n$ values, i.e. with a broad range of cluster lengths $\tau$ spanning more than one decades of values in the power law distribution,  a horizon dependence $H(M,n)$ varying with $M$ is found.
\par
For identically distributed sequences of clusters, $I(M,n)$ does not change with $M$ regardless of the value of $n$. This, has been shown in Fig. \ref{Fig:entropysampledpriceM1612Artif} where the cluster entropy $S (\tau,n)$ of artificially generated series (fractional random walks) are shown. One can note that the curves are practically unchanged at varying horizons $M$ and cluster duration $\tau$.
The departure from the {\em iid} case can be taken as a measure of price dynamics.
\par
Furthermore, by comparing the figures corresponding to the different assets a dependence of the function $I(n)$ is observed.   In the case of the NASDAQ the variation seems larger than for the S$\&$P500, and even larger than for the DJIA.


\section{Discussion and Conclusions}
\label{Discussion}
Next, the main results of the analysis of the \emph{cluster entropy} $S (\tau,n)$ and the \emph{cumulative information measure} $I(M,n)$  are compared with the results obtained by using information theoretical approaches by other authors.

\par
To build a cluster entropy index of horizon dependence, i.e. a synthetic numerical parameter with the ability to provide an estimate of the horizon dependence, we consider the entropy integral  $I(n)$ defined by Eq.~(\ref{Integrald}) at one-period  ($M=1$)  and at multiples of one period $M$ respectively defined as $I(1,n)$  and $I(M,n)$.   The quantity $I(M,n)$, defined above on the basis of  Eq.~(\ref{Integrald})   is called \emph{Market Dynamic Index}.
\par
\begin{figure}
\includegraphics[scale=0.29]{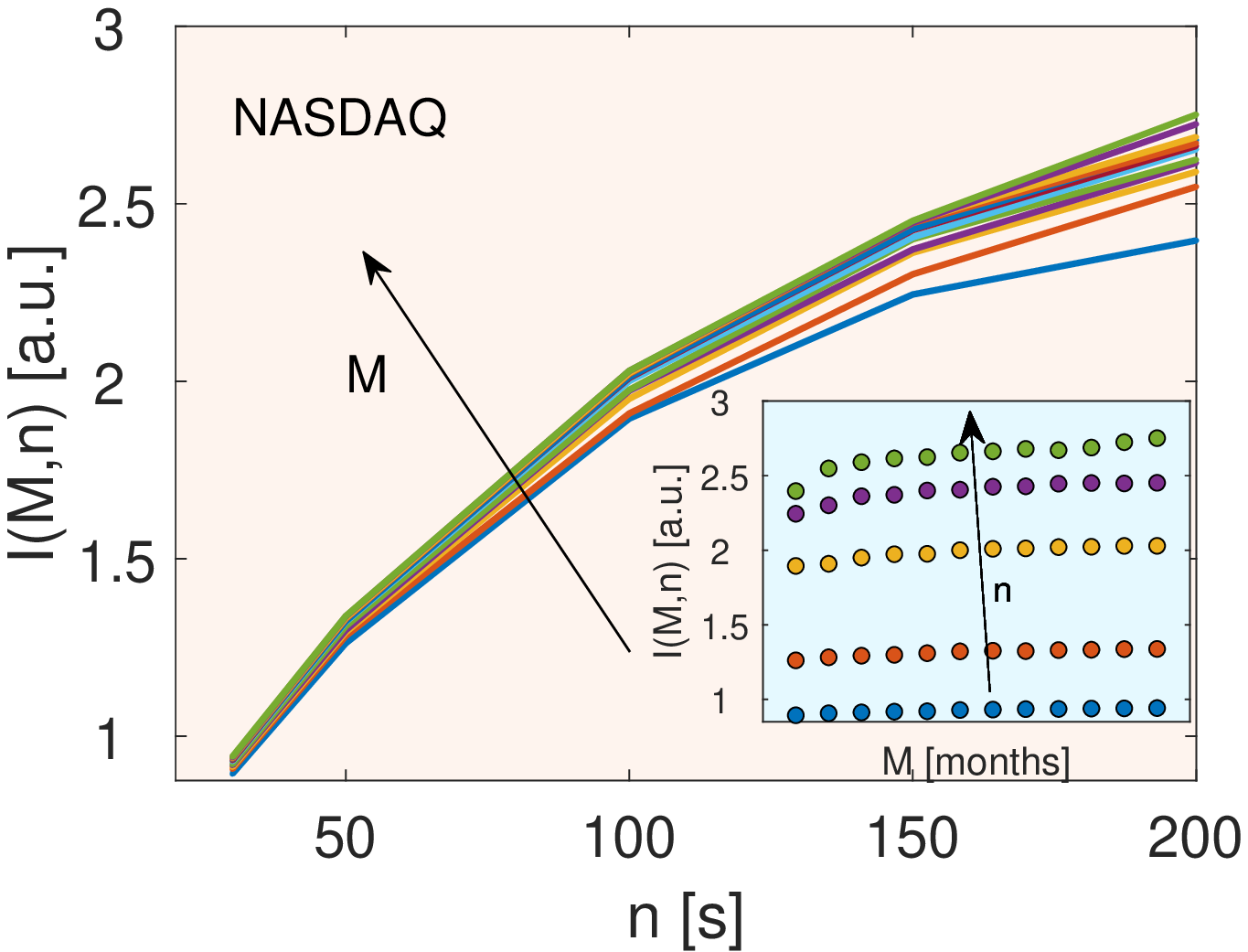}
\includegraphics[scale=0.29]{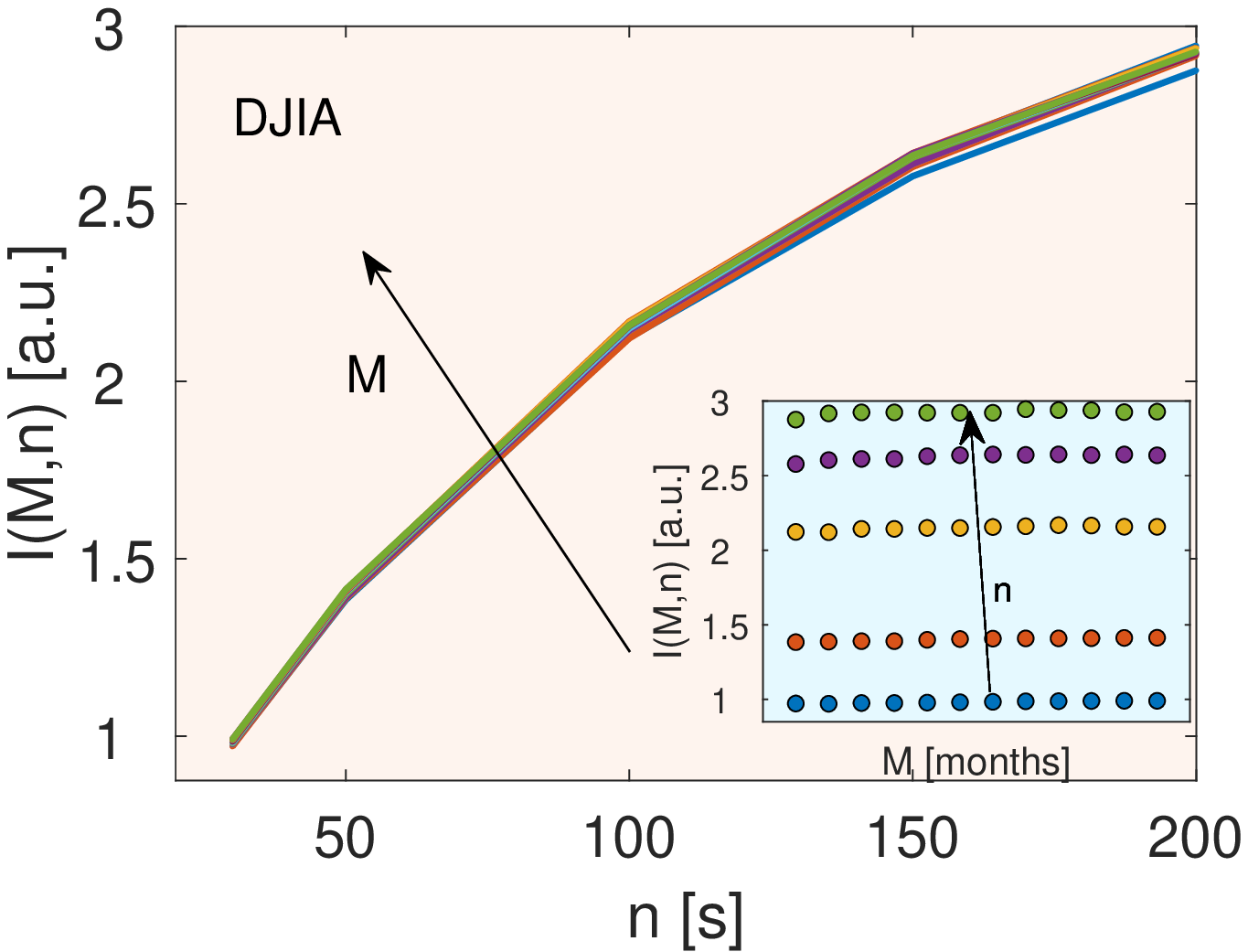}
\includegraphics[scale=0.29]{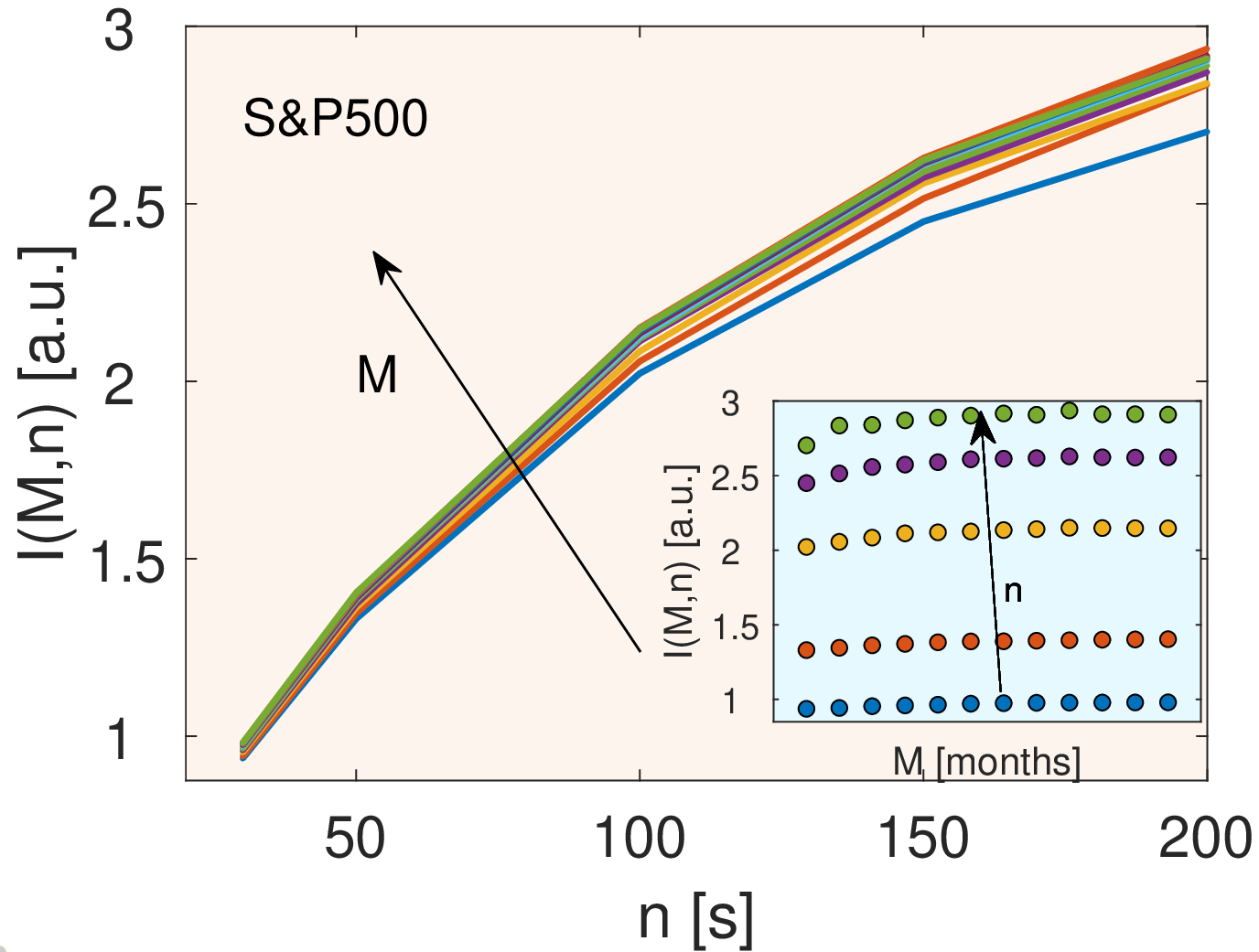}
\caption{\emph{Market Dynamic Index} $I(M,n)$ as a function of  the moving average window $n$, calculated according to Eq.~(\ref{Integral}) for the prices respectively of the Nasdaq Composite, Dow Jones Industrial Average and S$\&$P500 indexes as described in Table~\ref{tab:sampleddata}. Different curves in each figure refer to horizon varying from one ($M=1$) to twelve months ($M=12$) . In particular, this set of curves corresponds to time series length  $N=492035$ with sampling frequency calculated as described Section \ref{Data}.
$I(M,n)$ has been evaluated as the integral of the entropy curves $S(\tau, n)$ similar to those shown in Figure \ref{Fig:entropysampledpriceM1}.
The insets show \emph{ Market Dynamic Index} $I(M,n)$ as a function of the horizon period unit $M$. Symbols with different colors refer to different values of the moving average window $n$ as indicated by the arrow (namely $n = 30s$, $n = 50s$, $n = 100s$, $n = 150 s$ and $n = 200 s$).
}
\label{Fig:integral}
\end{figure}

To the purpose of comparing our results with those  of paper  \cite{backus2014sources},  the horizon dependence $H(M,n)$  is calculated as:
\begin{equation}
\label{horizon}
H(M,n) = I(M,n)-I(1,n) \hspace{5pt}.
\end{equation}
Values of \emph{Market Dynamic Index}  $I(M,n)$ and horizon dependence $H(M,n)$  calculated by using the NASDAQ, DJIA and S$\&$P500 data are reported in Table \ref{tab:horizonsNASDAQ}. The quantity $I(1,n)=I(1)$ is a reference value of the one-period entropy (lower bound). It is taken as  $I(1)=0.0049$, $I(1)=0.0214$ and $I(1)=0.0197$ respectively for power utility, recursive utility and difference habit agent models of the consumption growth following \cite{backus2014sources}. The value $I(12,n)$ has been obtained from the curves in Fig.~\ref{Fig:integral} for the prices of the NASDAQ, DJIA and S$\&$P500. $H(12,n)$  is the difference between $I(12,n)$ and $I(1,n)$ on account of Eq.~(\ref{horizon}).
\par
Next, the values of the horizon dependence obtained by using the cluster entropy  will be checked against those obtained by using different representative agent models for the definition of the pricing kernel in \cite{backus2014sources}.
The pricing kernel dynamics has been quantified by a measure of  entropy dependence on the investment horizon for popular asset pricing models. The pricing kernel accounts for the stochastic dynamic evolution of asset returns, which in their turn contain information about the pricing kernel. The analysis is based on the Kulback-Leibner divergence (also known as relative entropy) of the true probability distribution of the prices with respect to the risk-adjusted probability. On account of those results, it was argued that a realistic asset pricing model should have substantial one-period entropy and modest horizon dependence to justify   equity mean excess returns and bond yields at once.
\par
The  Kullback-Leibler (KL) divergence of the continuous probability measure $p(x)$ with respect
to some probability measure $p^*(x_t)$,  writes:
\begin{equation}
\label{KL1}
{KL}(P||P^*) = \int_X  p(x_t) \log \left( \frac{ p(x_t)}{p^*(x_t)}\right) dx_t
\end{equation}
Eq.~(\ref{KL1}) can be interpreted as the expectation of the function $\log { p(x_t)}/{p^*(x_t)}$ with respect to the probability $p(x_t)$:
\begin{equation}
\label{KL2}
{KL}(P||P^*)= E \left[\log \left( \frac{ p(x_t)}{p^*(x_t)}\right)\right]
\end{equation}
It can be easily shown that the relative entropy Eq.~(\ref{KL1}) reduces to Eq.~(\ref{Shannon_int}) for constant  probability $p^*(x_t)$.
\par
 Investigation of asset price dispersion and dynamics has been put forward by using a variant of the Kullback-Leibler (KL) divergence  of the pricing kernels $m_{t,t+n}$  expressed in terms of the ratio  between the true and risk-adjusted distribution \cite{backus2014sources}.   In this work, different representative agent models have been considered to quantify the {\em Market Horizon Dependence} $H(M)$ :
\begin{equation}
H(M)= I(M) -I(1)
\end{equation}
with the quantity $I(M)$  defined as:
\begin{equation}
I(M)= \frac{EL_t(m_{t,t+M})}{M} \hspace{10pt}.
\end{equation}
where ${EL_t(m_{t,t+M})}$ is defined as the average of the relative entropy of the pricing kernel,  and $I(1)$ is calculated at the month $M=1$. A summary of the horizon dependence  obtained by estimating the Kullback-Leibler (KL) entropy with pricing kernels generated by different representative agent models  according to the approach of \cite{backus2014sources} is reported in Table \ref{tab:hdconstant}.
\par
\begin{figure}[h]
\includegraphics[scale=0.3]{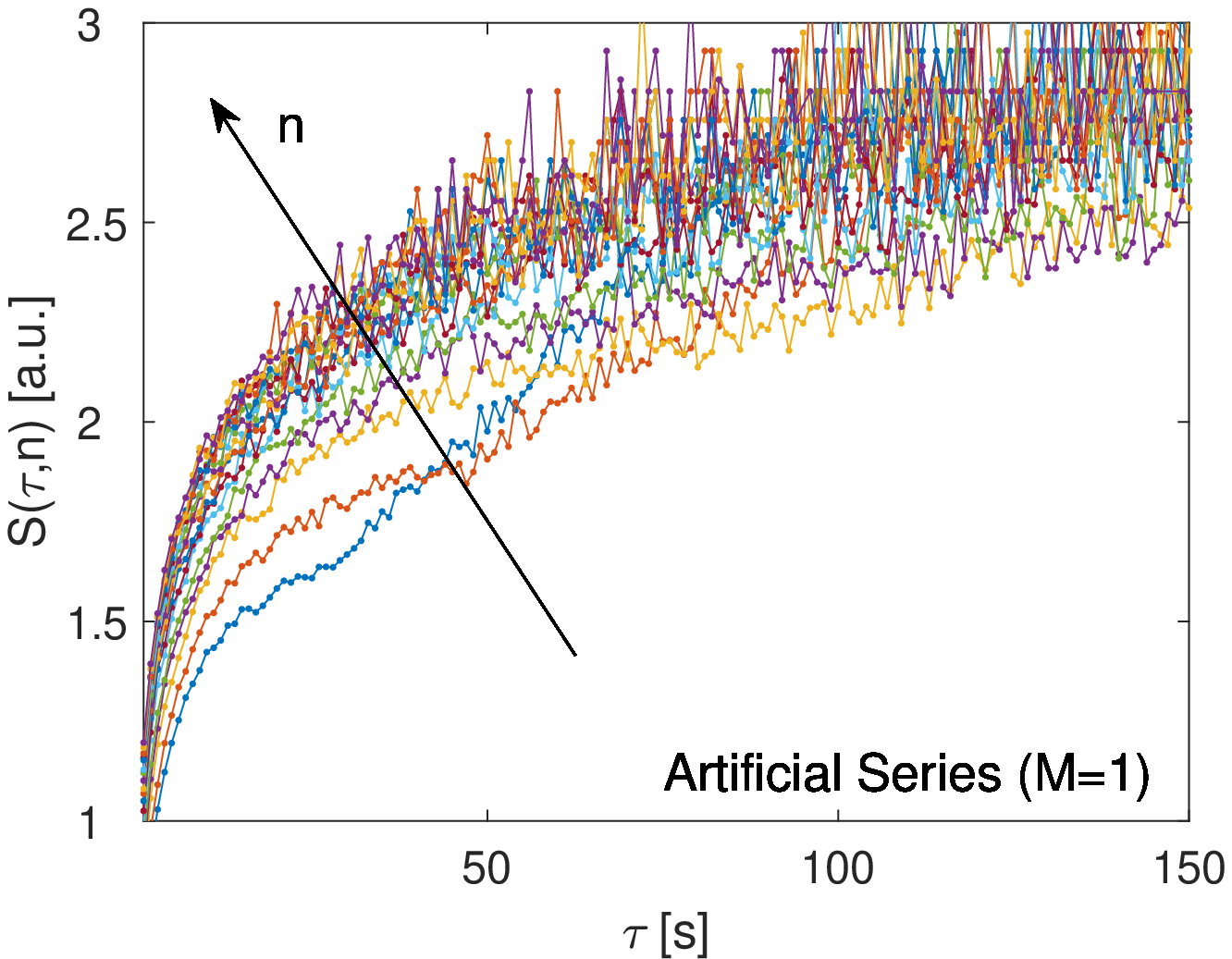}
\includegraphics[scale=0.3]{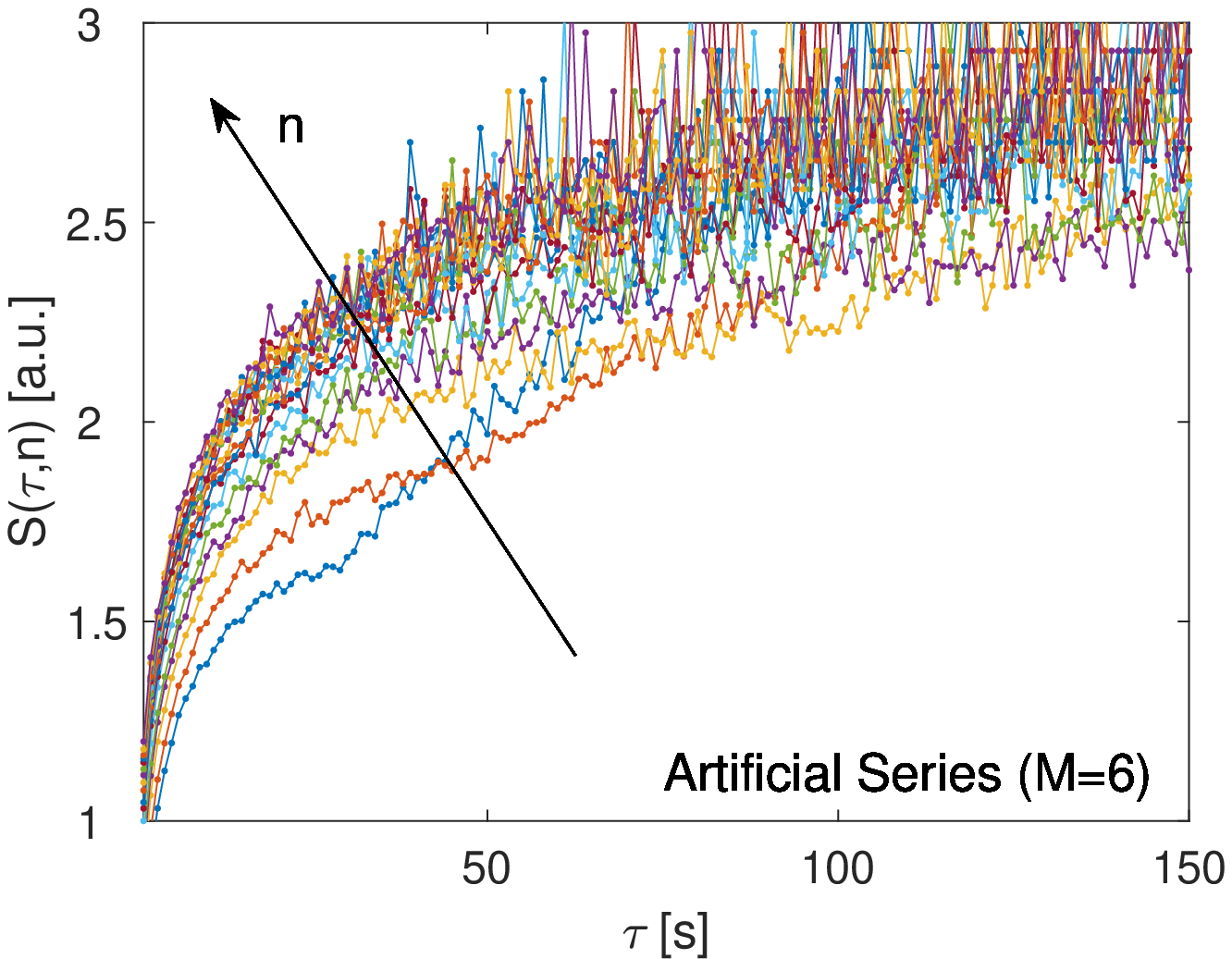}
\includegraphics[scale=0.3]{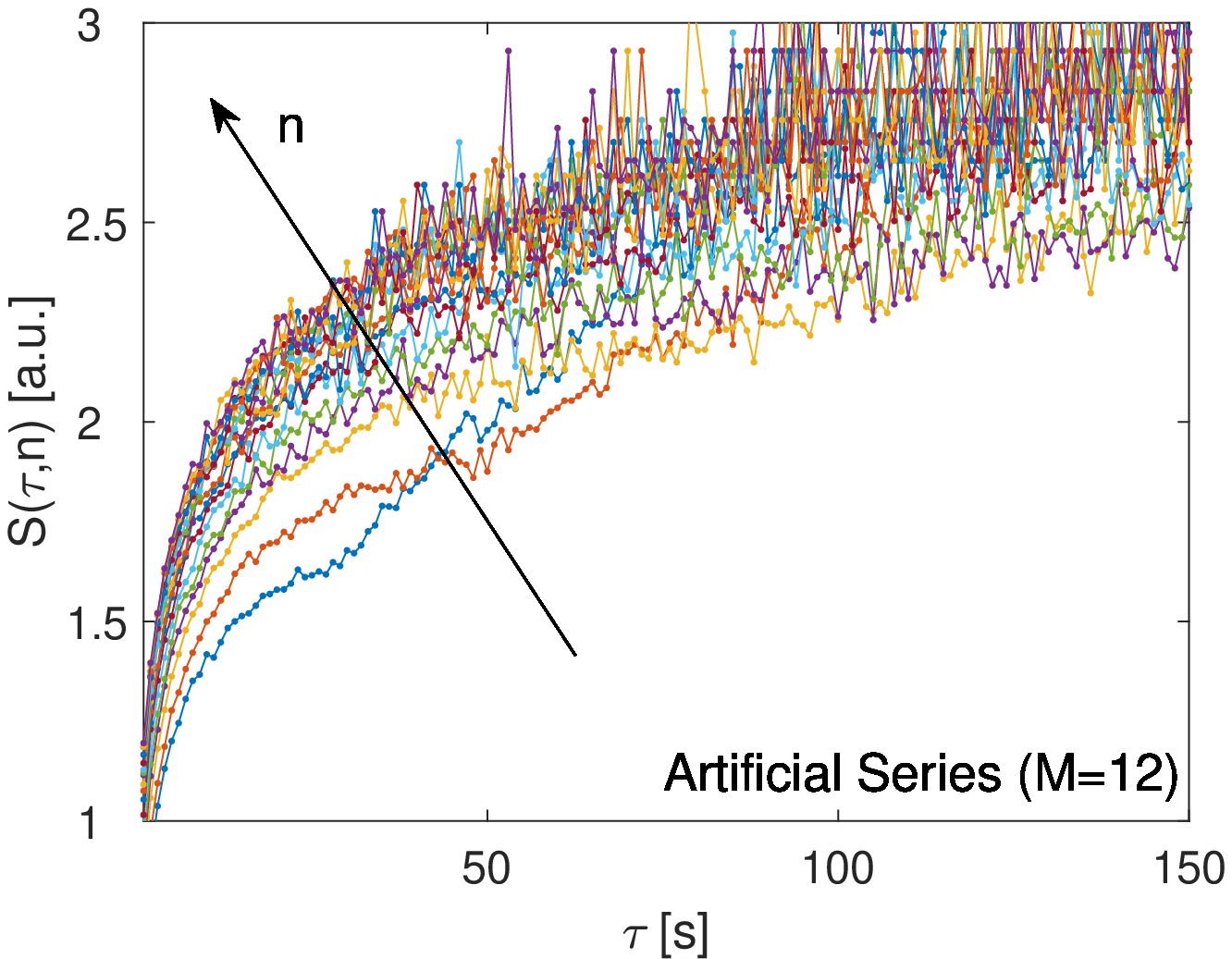}
   \caption{
   Cluster entropy $S(\tau,n)$  plotted   versus cluster duration $\tau$ for the time series of the prices (sampled data) of the market indices $H=0.5$.  Figures refer to one ($M=1$), six ($M=6$) and twelve months of data ($M=12$). All time series have same length $N=492035$ obtained by a suitable sampling frequency.
   Different curves refer to different values of the moving average window $n$.
      }
 \label{Fig:entropysampledpriceM1612Artif}
\end{figure}

To further validate the behaviour observed in real-world financial markets, simulations have been performed on artificial data generated by means of the FRACLAB tool available at:  https://project.inria.fr/fraclab/. The FRACLAB tool generates Fractional Brownian Motion series with assigned Hurst exponent $H$. The Hurst exponent corresponding to financial prices is generally assumed to be $H\sim 0.5$.  In Fig.~\ref{Fig:entropysampledpriceM1612Artif}, the cluster entropy curve is shown for a FBM series with $H=0.5$ and different lengths  $N$. For the curves shown in Fig.~\ref{Fig:entropysampledpriceM1612Artif} the artificial series has been generated with a total length equal to the one of the  NASDAQ index ($N=6982017$). Then the artificial series has been divided in twelve consecutive segments with the same lengths of the NASDAQ sub-sequences (values of first column of Table \ref{tab:sampleddata}). Figures refer respectively to the first segment ($M=1$), the first sixth segments ($M=6$) and  twelve segments ($M=12$).
\par
\begin{figure}[h]
\includegraphics[scale=0.29]{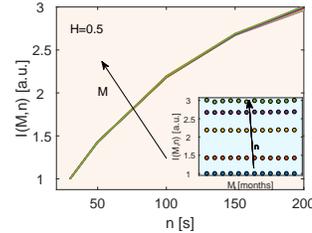}
\caption{\emph{Market Dynamic Index} $I(M,n)$ as a function of  the moving average window $n$, calculated according to Eq.~(\ref{Integral}) for the prices  $H=0.5$. Different curves in each figure refer to horizon varying from one ($M=1$) to twelve months ($M=12$) . In particular, this set of curves corresponds to time series length  $N=492035$ with sampling frequency calculated as described Sect. \ref{Data}.
$I(M,n)$ has been evaluated as the integral of the entropy curves $S(\tau, n)$ similar to those shown in Fig. \ref{Fig:entropysampledpriceM1}.
The insets show \emph{ Market Dynamic Index} $I(M,n)$ as a function of the horizon period unit $M$. Symbols with different colors refer to different values of the moving average window $n$ as indicated by the arrow (namely $n = 30s$, $n = 50s$, $n = 100s$, $n = 150 s$ and $n = 200 s$).
}
\label{Fig:integralMDIArtif}
\end{figure}

To the purpose of fully appreciating the different  behaviour of real world market series compared to those exhibited by the artificially generated sequences, the market dynamic index  has been calculated for the artificial series  (Fig. \ref{Fig:integralMDIArtif}). The Market Dynamic Index has quite a constant value at the different horizons $M$ and moving average clusters $n$, thus exhibiting a behaviour different than real market dynamic indexes shown in Fig.~\ref{Fig:integral}.
\par
Last but not least, results of statistical significance tests are reported in Table \ref{tab:significance}. The test has been performed by using  the paired t-test to check the null hypothesis $h=0$ that the cluster entropy values obtained on the real-world financial markets and those obtained on the artificial series (FBMs with $H=0.5$ assumed as benchmark) come from distributions with equal mean and same variance with a probability $p$.
\par
One can note in Table \ref{tab:significance} that the probability $p$  that the null hypothesis holds true ranges from $0.5154$ to $0.7584$. This confirms  that the NASDAQ market behaves quite differently from the traditional interpretation of independent elementary stochastic process of price variations, as the FBM with $H=0.5$ implies.
The S\&P 500 exhibits an intermediate tendency to behave as ideal market being the probability $0.7399\leq p \leq 0.9248$. The probability for Dow Jones ranges within the interval $0.8892\leq p \leq 0.9434$. Thus it seems that the DJIA index reproduces more closely the behaviour of the fully independent stochastic process involved in the FBM with $H=0.5$.
\par
From an economic perspective the results have shown how financial market apparently very similar in terms of regional features, size and volumes may exhibit different horizon dependence. The obtained results are very robust from a statistical point of view. Thus they can represent a valid basis for developing  investment tools  to  quantify risks and extremely useful for investors in classifying markets and choosing their strategy.
\clearpage
\newpage

\clearpage
\newpage
\begin{table}[h]
\centering
\scriptsize
\setlength{\tabcolsep}{0.1cm}
\renewcommand{\arraystretch}{1.2}
\caption{\emph{Markets}. NASDAQ, DJIA and S$\&$P500  data sets have been downloaded from the terminal  www.bloomberg.com/professional. The analysis has been carried on tick-by-tick data of US indexes traded over the year 2018.  Tick duration (time interval between individual transactions) is about one second for all the three markets.}
\label{tab:data}
\begin{tabular}{|c|p{4cm}|c|c|c|c|}
\hline
\rowcolor{purple!40!orange!40}
Ticker & Name &            Country              & Currency       & Members & Length  \\
\hline
NASDAQ & Nasdaq Composite     & US & USD & 2570 & 6982017 \\
DJIA & Dow Jones Industrial Average &        US & USD &       30 & 5749145 \\
S$\&$P500 &    Standard \& Poor 500  & US & USD & 505 & 6142443 \\
\hline
\end{tabular}
\end{table}

\bigskip
\bigskip
\bigskip
\begin{table}[h]
\centering
\scriptsize
\setlength{\tabcolsep}{0.1cm}
\renewcommand{\arraystretch}{1.2}
\caption{\emph{Data Length}. First column reports the temporal horizon $M$ (number of periods in month units). Second, third and fourth columns report the length $N$ of the price series for each temporal horizon $M$ for the three assets. }
\label{tab:sampleddata}
\begin{tabular}{|c|c|c|c|}
\hline
\rowcolor{purple!40!orange!40}
 \multicolumn{4}{|c|}{Data Length $N$}  \\
\hline
\rowcolor{purple!10!orange!10}
$M$   & NASDAQ &   DJIA & S$\&$P500   \\
\hline
1  & 586866 &   516644 &           516635 \\
2  & 1117840 & 984101 &           984046  \\
3  & 1704706 & 1500764             & 1500662 \\
4  & 2291572 & 1623779 &         2017282 \\
5  & 2906384     & 2165044 &     2558504 \\
6 & 3493250      & 2681708 &     3075125 \\
7  & 4069315 & 3187571 &         3580946 \\
8  & 4712062 & 3753440             & 4146769 \\
9   & 5243029    & 4220774 &     4614186\\
10 & 5885781    & 4786624         & 5180006  \\
11  & 6461845   & 5292487 &     5685826 \\
12  & 6982017 &            5749145 &         6142443  \\
\hline
\end{tabular}
\end{table}
\clearpage
\newpage
\begin{table}[h]
\centering
\scriptsize
\setlength{\tabcolsep}{0.1cm}
\renewcommand{\arraystretch}{1.2}
\caption{\emph{Market Dynamic Index} $I(M,n)$ and \emph{Market Horizon Dependence} $H(M,n)$.  The indexes  $I(M,n)$ and $H(M,n)$ have been calculated by using the relationships Eqs.(\ref{Integrald},\ref{horizon}) with entries the entropy data like those plotted in Figs. (\ref{Fig:entropysampledpriceM1},
\ref{Fig:entropysampledpriceM12}) for the NASDAQ, DJIA and S$\&$P500 indexes. The spanned horizons $M$ range from one  to twelve months (respectively from $M=1$ to $M=12$). The values of the moving average window $n$  are reported in the first column. The reference values $I(1)$ in the third, fourth and fifth column have been taken equal to those of the consumption growth models respectively with power utility ($I(1)=0.0049$), recursive utility ($I(1)=0.0214$) and difference habit ($I(1)=0.0197$). 
}
\label{tab:horizonsNASDAQ}
\begin{tabular}{|c|c|c|c|c|}
\hline
\rowcolor{purple!40!orange!40}
\multicolumn{5}{|c|}{ Nasdaq Composite Index (NASDAQ) } \\
\hline
\rowcolor{purple!10!orange!10}
         $n$  & Entropy Indexes & Power Utility &      Recursive Utility              & Difference Habit   \\
\hline
{30} & ${I(12)}$ & {0.0052} & {0.0226}  & {0.0208}    \\
& ${H(12)}$ & {0.0003} & {0.0012} & {0.0011}  \\
\hline
{50} & ${I(12)}$ & {0.0052} & {0.0227}  & {0.0209}    \\
& ${H(12)}$ & {0.0003} & {0.0013} & {0.0012}   \\
\hline
{100} & ${I(12)}$ & {0.0052} & {0.0229}  & {0.0211} \\
& ${H(12)}$ & {0.0003} & {0.0015} & {0.0014}   \\
\hline
{150} & ${I(12)}$ & {0.0054} & {0.0234}  & {0.0215}      \\
& ${H(12)}$ & {0.0005} & {0.0020} & {0.0018}  \\
\hline
{200} & ${I(12)}$ & {0.0056} & {0.0246}  & {0.0226}     \\
& ${H(12)}$ & {0.0007} & {0.0032} & {0.0029}   \\
\hline
\hline
\rowcolor{purple!40!orange!40}
\multicolumn{5}{|c|}{Dow Jones Industrial Average Index (DJIA) } \\
\hline
\rowcolor{purple!10!orange!10}
         $n$  & Entropy Indexes & Power Utility &      Recursive Utility              & Difference Habit   \\
\hline
{30}& $I(12)$  & 0.0050    & 0.0218  & 0.0201     \\
& $H(12)$  & 0.0001 & 0.0004 & 0.0004   \\
\hline
{50}& $I(12)$  & 0.0050    & 0.0219  & 0.0201    \\
& $H(12)$  & 0.0001  & 0.0005& 0.0004   \\
\hline
{100}& $I(12)$ & 0.0050    & 0.0217  & 0.0200    \\
& $H(12)$ & 0.0001 & 0.0003& 0.0003  \\
\hline
{150}& $I(12)$ & 0.0050    & 0.0219  & 0.0201     \\
& $H(12)$ & 0.0001  & 0.0005 &0.0004  \\
\hline
{200}& $I(12)$ & 0.0050    & 0.0218  & 0.0201     \\
& $H(12)$ & 0.0001 & 0.0004 & 0.0004 \\
\hline
\hline
\rowcolor{purple!40!orange!40}
\multicolumn{5}{|c|}{ S\&P 500 Index (S$\&$P500) } \\
\hline
\rowcolor{purple!10!orange!10}
         $n$  & Entropy Indexes & Power Utility &      Recursive Utility              & Difference Habit   \\
\hline
{30} & ${I(12)}$ & {0.0051} & {0.0224}  & {0.0206}     \\
& ${H(12)}$ & {0.0002} & {0.0010} & {0.0009}    \\
\hline
{50} & ${I(12)}$ & {0.0052} & {0.0226}  & {0.0208}    \\
& ${H(12)}$ & {0.0003} & {0.0012} & {0.0011}    \\
\hline
{100} & ${I(12)}$ & {0.0052} & {0.0227}  & {0.0209}   \\
& ${H(12)}$ & {0.0003} & {0.0013} & {0.0012}  \\
\hline
{150} & ${I(12)}$ & {0.0052} & {0.0229}  & {0.0211}     \\
& ${H(12)}$ & {0.0003} & {0.0015} & {0.0014}   \\
\hline
{200} & ${I(12)}$ & {0.0053} & {0.0230}  & {0.0212}     \\
& ${H(12)}$ & {0.0004} & {0.0016} & {0.0015}  \\
\hline
\end{tabular}
\end{table}
\clearpage
\newpage
\begin{table}[h]
\centering
\scriptsize
\setlength{\tabcolsep}{0.1cm}
\renewcommand{\arraystretch}{1.2}
\caption{\emph{Entropy Index and Horizon Dependence}. The indexes are obtained by using a Kullback-Leibler entropy approach for  representative agent models with constant variance (top), stochastic variance (middle) and jumps (bottom) as estimated by \cite{backus2014sources}.}
\label{tab:hdconstant}
\begin{tabular}{|c|c|c|c|c|}
\hline
\rowcolor{purple!40!orange!40}
\multicolumn{5}{|c|}{Constant Variance } \\
\hline
\rowcolor{purple!10!orange!10}
Entropy & Power Utility &     Recursive Utility              & Ratio Habit   & Difference Habit  \\
\hline
${I(1)=E L_{t}\left(m_{t, t+1}\right)}$ & {0.0049} & {0.0214} & {0.0049} & {0.0197}  \\
${I(\infty)}$ & {0.0258} & {0.0232} & {0.0003} & {0.0258}   \\
${H(120)=I(120)-I(1)}$ & {0.0119} & {0.0011} & {-0.0042} & {0.0001}  \\
${H(\infty)=I(\infty)-I(1)}$ & {0.0208} & {0.0018} & {-0.0047} & {0.0061}   \\
\hline
\hline
\rowcolor{purple!40!orange!40}
\multicolumn{5}{|c|}{Stochastic Variance} \\
\hline
\rowcolor{purple!10!orange!10}
Entropy &     Recursive Utility 1          & Recursive Utility 2     & Campbell Cochrane & -\\
\hline
${I(1)=E L_{t}\left(m_{t, t+1}\right)}$ & {0.0218} & {0.0249} & {0.0230} &  -\\
${I(\infty)}$ & {0.0238} & {0.0293} & {0.0230}&   - \\
${H(120)=I(120)-I(1)}$ & {0.0012} & {0.0014} & {0} & - \\
${H(\infty)=I(\infty)-I(1)}$ & {0.0020} & {0.0044} & {0} & -  \\
\hline
\hline
\rowcolor{purple!40!orange!40}
\multicolumn{5}{|c|}{with Jumps} \\
\hline
\rowcolor{purple!10!orange!10}
Entropy & IID w/Jumps &      Stochastic Intensity        & Constant Intensity 1              & Constant Intensity 2  \\
\hline
${I(1)=E L_{t}\left(m_{t, t+1}\right)}$ & {0.0485} & {0.0512} & {1.2299} & {0.0193}  \\
${I(\infty)}$ & {0.0485} & {0.0542} & {15.730} & {0.0200}    \\
${H(120)=I(120)-I(1)}$ & {0} & {0.0025} & {9.0900} & {0.0005}  \\
${H(\infty)=I(\infty)-I(1)}$ & {0} & {0.0030} & {14.5000} & {0.0007}   \\
\hline
\end{tabular}
\end{table}

\bigskip
\bigskip
\bigskip

\begin{table}[h]
\centering
\scriptsize
\setlength{\tabcolsep}{0.1cm}
\renewcommand{\arraystretch}{1.2}
\caption{\emph{T-paired test}. First column reports the temporal horizon $M$. Second, third and fourth column report the probability $p$ to reject the null hypothesis, which corresponds to assume that the cluster entropy values (for the NASDAQ, DJIA and S$\&$P500 at varying horizons $M$) has same mean and variance of the Fractional Brownian Motion with $H=0.5$.}
\label{tab:significance}
\begin{tabular}{|c|c|c|c|}
\hline
\rowcolor{purple!40!orange!40}
\multicolumn{4}{|c|}{Probability $p$}  \\
\hline
\rowcolor{purple!10!orange!10}
$M$   & NASDAQ &   DJIA & S$\&$P500   \\
\hline
1  & 0.5154	& 0.8892	& 0.7399
 \\
2  & 0.6026	& 0.9257	& 0.8335
  \\
3  & 0.6470	& 0.9332	& 0.8588
\\
4  & 0.6631 & 0.9283	& 0.8814
 \\
5  & 0.6823 &	0.9417	 & 0.9018
 \\
6 & 0.7124 & 0.9534	& 0.9246
 \\
7  & 0.7162 & 0.9461	& 0.9224
 \\
8  & 0.7288 & 0.9618 &	0.9309
\\
9   & 0.7370 & 0.9645	& 0.9479
\\
10 & 0.7409 & 0.9570 &	0.9336
 \\
11  &0.7542 &	0.9519	& 0.9321
\\
12  & 0.7584	& 0.9434	& 0.9248
  \\
\hline
\end{tabular}
\end{table}
\clearpage
\newpage

\newpage
\clearpage

\clearpage
\newpage

\clearpage
\newpage
\section*{Supporting information}

This section describes the files available in the Dropbox Shared Folder:$ https://www.dropbox.com/sh/9pfeltf2ks0ewjl/AACjuScK_gZxmyQ_mDFmGHoya?dl=0$
\par
In particular the following items can be found in the dropbox folder:
\begin{itemize}
  \item Data folder
  \item Codes zipped file
\end{itemize}

\subsection*{DATA}
H05Art....zip refers to two data folders. These data can be generated and analysed by using the codes described here below.
Horizon.zip contains the horizon (.MAT) data estimated for market series and artificial series shown in Figures 5 and 7. These data have been used to estimate the horizon dependence reported in Table 3.  These data have been used to perform the test whose results are reported in Table 5.

\subsection*{CODES}
Codes.zip: contains all the MATLAB  codes used for the analysis.\\
The following table provides further details regarding each file included in the Code zipped folder and the algorithm steps.

\begin{table}[h]
\centering
\scriptsize
\setlength{\tabcolsep}{0.1cm}
\renewcommand{\arraystretch}{1.2}
\caption{List of .m files included in Code.zip}
\label{tab:listfile}
\begin{tabular}{|c|p{9.0cm}|}
\hline
\rowcolor{purple!40!orange!40}
\multicolumn{2}{|c|}{List of .m files included in Code.zip}  \\
\hline
\rowcolor{purple!10!orange!10}
FileName  & Description  \\
\hline
TimeSeriesGenerator.m  & generates time series using the FracLab matlab environment. \\
\hline
SampledSeries.m  & First, creates the data structures. Second it samples the data. It includes the variable “Sampled”. If “Sampled=1” the output data are sampled. If “Sampled=0” the data are raw (i.e. unsampled).  \\
\hline
Prices.m & It evaluates and saves the prices vector under analysis\\
\hline
Entropy.m  & evaluates the cluster entropy of the series. It generates the curves plotted in Figures 1, 2, 3, 4 and 6 in the paper. \\
\hline
MarketDynamicIndex.m  & evaluates the Market Dynamic Index according to Eq. (12).\\
\hline
DMIfigure.m & Plots the Market Dynamic Index, i.e. it yields the Figures 5 and 7 in the paper. \\
\hline
DMA.m & Function \\
\hline
DMAbackward.m  & Function \\
\hline
DMAcentered.m   & Function  \\
\hline
DMAforward.m & Function    \\
\hline
ComputeClusterProbability.m  & Function  \\
\hline
\end{tabular}
\end{table}

\clearpage
In Table \ref{tab:workflowfile} the main calculation steps are summarized:
\begin{itemize}
  \item Generate the data with the \textit{“TimeSeriesGenerator.m”} file or use already available timeseries in the specified format.
  \item Create a structured data with option sampled/unsampled the series with the \textit{“SampledSeries.m”} file.
  \item Create a vector of prices by means of  \textit{“Prices.m”}
  \item Evaluate the entropy by means of \textit{“Entropy.m”}  (to produce Figure 1,2 for unsampled data series; to produce Figure 3 and 4 for sampled data series) .
  \item Evaluate the Market Dynamic Index by means of \textit{“MarketDynamicIndex.m”}
  \item Plot the MarketDynamic Index by means of \textit{“MDI$\_$figure.m"}  (Figure 5 and Figure 7).
\end{itemize}

\begin{table}[h]
\centering
\scriptsize
\setlength{\tabcolsep}{0.1cm}
\renewcommand{\arraystretch}{1.2}
\caption{Workflow of .m files included in Code.zip}
\label{tab:workflowfile}
\begin{tabular}{|c|p{5cm}|p{4.5cm}|}
\hline
\rowcolor{purple!40!orange!40}
\multicolumn{3}{|c|}{Workflow of .m files included in Code.zip}  \\
\hline
\rowcolor{purple!10!orange!10}
FileName  & Input files (Data and Functions) & Output files (Data)  \\
\hline
TimeSeriesGenerator.m  &   None &          Data1.mat, $\ldots$ , Data12.mat \\
\hline
SampledSeries.m  & Data1.mat, $\ldots$, Data12.mat & DataSampled0.mat; \hspace{15pt} DataSampled1.mat  \\
\hline
Prices.m & DataSampled0.mat, DataSampled1.mat             & PricesData1.mat, $\ldots$, PricesData12.mat. \\
\hline
Entropy.m  & PricesData1.mat, $\ldots$, PricesData12.mat; DMA.m; ComputeClusterProbability.m &         Simulations$\_$Complete$\_$Data1.mat,$\ldots,$ Simulations$\_$Complete$\_$Data12.mat \\
\hline
MarketDynamicIndex.m  &  Simulations$\_$Complete$\_$Data1.mat, $\ldots,$ Simulations$\_$Complete$\_$Data12.mat &     MDI$\_$n$\_$Prices$\_$Data.mat \\
\hline
DMI$\_$figure.m & MDI$\_$n$\_$Prices$\_$Data.mat &  None    \\
\hline
DMA.m & DMA$\_$backward.m; DMA$\_$centered.m; DMA$\_$forward.m &       None \\
\hline
DMA$\_$backward.m  & None             & None \\
\hline
DMA$\_$centered.m   &  None &     None\\
\hline
DMA$\_$forward.m &  None         & None  \\
\hline
ComputeClusterProbability.m  & None &    None \\
\hline
\end{tabular}
\end{table}


{\textbf{Data Availability Statement}} \\
\par
Data of NASDAQ, DIJA, SP500 indexes have been downloaded from the Bloomberg terminal: www.bloomberg.com/professional, which is accessible to several organizations worldwide or by activating a subscription via the registration page. \\
\par
Furthermore we provide a set of series of different lengths (simulating the different horizons) that have been generated artificially by means of the FRACLAB tools freely accessed at:  https://project.inria.fr/fraclab/ and a MATLAB code to reproduce the results shown in the Figures.


\end{document}